\documentclass[10pt,conference]{IEEEtran}
\usepackage{cite}
\usepackage{amsmath,amssymb,amsfonts}
\usepackage{graphicx}
\usepackage{textcomp}
\usepackage{xcolor}
\usepackage[hyphens]{url}
\usepackage{fancyhdr}
\usepackage{hyperref}

\usepackage{comment}
\usepackage{algorithm}
\usepackage{algpseudocode}
\usepackage{multirow}
\usepackage{tablefootnote}
\usepackage{makecell}
\usepackage{array}
\usepackage{pifont}
\newcommand{\ignore}[1]{}
\usepackage{threeparttable}
\usepackage{float}
\usepackage[normalem]{ulem}   % \sout for strikethrough
\usepackage{enumitem}
\usepackage{tcolorbox}
\usepackage{tabularx}
\usepackage[numbers, sort&compress]{natbib}
\usepackage{dblfloatfix}
\usepackage{soul}
\sethlcolor{yellow!30}

\usepackage{tikz}
\newcommand*\circled[1]{\tikz[baseline=(char.base)]{
            \node[shape=circle,draw,inner sep=0.5pt,fill=purple,text=white] (char) {#1};}}

\newcommand{\hpcayear}{2027}
\newcommand{\hpcasubmissionnumber}{1109}
\newcommand{\Dynamite}{DynaMite}
\newcommand{\review}[1]{#1}
\newcommand{\cmark}{{\color{green!70!black}\ding{51}}}
\definecolor{lblue}{HTML}{F6FBFC}
\newcommand{\bluebox}[1]{
\begin{tcolorbox}[colback=lblue,colframe=blue]
#1
\end{tcolorbox}
}

\title{\Dynamite{}: Dynamic Measurement Scheduling\\To Accelerate Fault-Tolerant Quantum Programs}
\def\hpcacameraready{} % Uncomment to build camera-ready version

\newcommand\hpcaauthors{Dongwhee Kim$^{\dagger}$, Suhas Vittal$^{\ddagger}$, and Poulami Das$^{\dagger}$}
\newcommand\hpcaaffiliation{$^{\dagger}$University of Texas at Austin, $^{\ddagger}$Georgia Institute of Technology}
\newcommand\hpcaemail{dongwhee.kim@utexas.edu, suhaskvittal@gatech.edu, poulami.das@utexas.edu}

\author{
  \ifdefined\hpcacameraready
    \IEEEauthorblockN{\hpcaauthors{}}
      \IEEEauthorblockA{
        \hpcaaffiliation{} \\
        \hpcaemail{}
      }
  \else
    \IEEEauthorblockN{\normalsize{HPCA \hpcayear{} Submission
      \textbf{\#\hpcasubmissionnumber{}}} \\
      \IEEEauthorblockA{
        Confidential Draft \\
        Do NOT Distribute!!
      }
    }
  \fi 
}

\fancypagestyle{camerareadyfirstpage}{%
  \fancyhead{}
  
  \fancyhead[C]{
    \ifdefined\aeopen
    \parbox[][12mm][t]{13.5cm}{\hpcayear{} IEEE International Symposium on High-Performance Computer Architecture (HPCA)}    
    \else
      \ifdefined\aereviewed
      \parbox[][12mm][t]{13.5cm}{\hpcayear{} IEEE International Symposium on High-Performance Computer Architecture (HPCA)}
      \else
      \ifdefined\aereproduced
      \parbox[][12mm][t]{13.5cm}{\hpcayear{} IEEE International Symposium on High-Performance Computer Architecture (HPCA)}
      \else
      \parbox[][0mm][t]{13.5cm}{\hpcayear{} IEEE International Symposium on High-Performance Computer Architecture (HPCA)}
    \fi 
    \fi 
    \fi 
    \ifdefined\aeopen 
      \includegraphics[width=12mm,height=12mm]{ae-badges/open-research-objects.pdf}
    \fi 
    \ifdefined\aereviewed
      \includegraphics[width=12mm,height=12mm]{ae-badges/research-objects-reviewed.pdf}
    \fi 
    \ifdefined\aereproduced
      \includegraphics[width=12mm,height=12mm]{ae-badges/results-reproduced.pdf}
    \fi
  }
  \fancyfoot[C]{}
}
\begin{document}
\maketitle

%Enables the camera ready header and footer
\ifdefined\hpcacameraready 
  \thispagestyle{camerareadyfirstpage}
  \pagestyle{empty}
\else
  \thispagestyle{plain}
  \pagestyle{plain}
\fi

\newcommand{\hpcaheight}{0mm}
\ifdefined\eaopen
\renewcommand{\hpcaheight}{12mm}
\fi

\thispagestyle{plain}\pagestyle{plain}   % arXiv: drop HPCA header
%%%%%%%%%%%%%%%%%%%%%%%%%%%%%%%%%%%%%%%%
%%%%%%%% -- PAPER CONTENT STARTS -- %%%%%%%%%

\begin{abstract}
Fault-tolerant Quantum Computers (FTQCs) run millions of logical cycles %during program execution 
to perform quantum error correction (QEC). As logical operations span entire logical cycles, the logical cycle time (LCT) dictates program runtime. Qubit measurements, critical to extract parity information for QEC, are the slowest operations and account for 43--86\% of LCT on state-of-the-art systems. Measurement latencies impact the operational fidelity--
fast measurements lower our ability to correctly classify quantum states, whereas slow measurements increase the likelihood of decoherence.
So, hardware providers calibrate measurement latencies that achieve a sweet spot and maximize readout fidelity per qubit. These latencies are then used to schedule \textit{all} measurements during program execution.
However, calibrated latencies are sub-optimal for logical performance because logical cycles span multiple QEC rounds, during which data qubits are not measured. %Our studies show that calibrated readout latencies are often conservative, which 
This increases LCT and logical error rates (LERs) due to a higher probability of decoherence at the logical level.

We propose \emph{\Dynamite{}}, a lightweight real-time measurement latency selection policy built on two key insights.
\textit{First}, logical cycles must use a measurement latency that is \textit{fast enough} to reduce both LCT and LER for the QEC code, instead of the calibrated latency that only maximizes readout fidelities.
\textit{Second}, not all logical cycles are equal; a large fraction of them often do not fully utilize the available \textit{error budget} of the QEC code, or the number of failed parity checks it can tolerate without compromising LER. These cycles could use even faster measurements but do not, because latencies are chosen statically prior to scheduling, whereas error information is only available at runtime.
\Dynamite{} exploits the fact that a logical cycle comprises multiple QEC rounds and selects the fastest latency for an upcoming round based on the error budget left and rounds remaining. 
Our evaluations show that \Dynamite{} reduces LCT by 1.69$\times$ on average and up to 2.29$\times$ and reduces LER to 0.46$\times$ on average and up to 0.13$\times$ of baseline.

\end{abstract}

\section{Introduction}

Quantum computers promise massive speedups for important problems in critical domains, such as chemistry, physics, cryptanalysis, etc~\cite{shor1999_polynomial, reiher2017elucidating, harrow2009_hhl, childs2018_first_qsim}. However, noisy hardware limits us from running most programs. \textit{Quantum error correction (QEC)} codes enable fault-tolerance by encoding logical qubits using redundant physical qubits~\cite{breuckmann2017hyperbolic, chamberland2020topological, google2021exponential, dennis2001_surface_codes, fowler2012surface, gottesman1997stabilizer, kitaev1997quantum, landahl2011fault, shor1995scheme, stein2025hetec}. The logical error rate (LER) decreases exponentially with increasing redundancy (or \textit{code distance}), if the error rate of physical qubits is below a \textit{threshold}~\cite{stephens2014fault, dennis2001_surface_codes}. QEC codes use data qubits to store program information and parity qubits to extract information about errors by using a \textit{syndrome extraction circuit}. The syndrome is used by a \textit{decoder} to correct errors.
Surface codes~\cite{kitaev2003fault, dennis2001_surface_codes, fowler2012surface, gidney2025yoked} are a leading QEC code due to high threshold (${\sim}1\%$) \cite{stephens2014fault}, availability of accurate and fast decoders~\cite{Higgott2025sparseblossom, das2022lilliput, delfosse2021almost, wu2023fusion, ravi2023better}, and nearest-neighbor connectivity needs. Recent real system studies~\cite{google2021exponential, google2025quantum, google2023suppressing} show their ability to exponentially suppress errors, solidifying their inclusion in industrial roadmaps~\cite {google_quantum_roadmap, fujitsu_quantum_strategy_2025}.
% IBM: qLDPC Code (\cite{ibm_quantum_roadmap})

Syndrome extraction is prone to gate and measurement errors on real systems. To tolerate these errors, decoders use `$d$' consecutive rounds of syndrome for a distance-$d$ code, which constitutes a \textit{logical cycle}.
\textit{Fault-Tolerant Quantum Computers (FTQC)} run programs by mapping program variables into logical qubits and translating instructions into logical operations. FTQCs interleave logical operations between logical cycles and only a single logical operation can be executed on a program qubit per logical cycle. So, the logical cycle time (LCT) dictates the throughput and runtime of fault-tolerant quantum programs~\cite{tannu2017taming, maurya2025_decoder_sync, gidney2021_rsa_factorization, mcardle2025_fast_for_the_curious}.
\textit{Consequently, reducing LCTs directly translates to reduced execution times and speedups.}

\begin{figure*}[!t]
  \centering                                 \includegraphics[width=\textwidth]{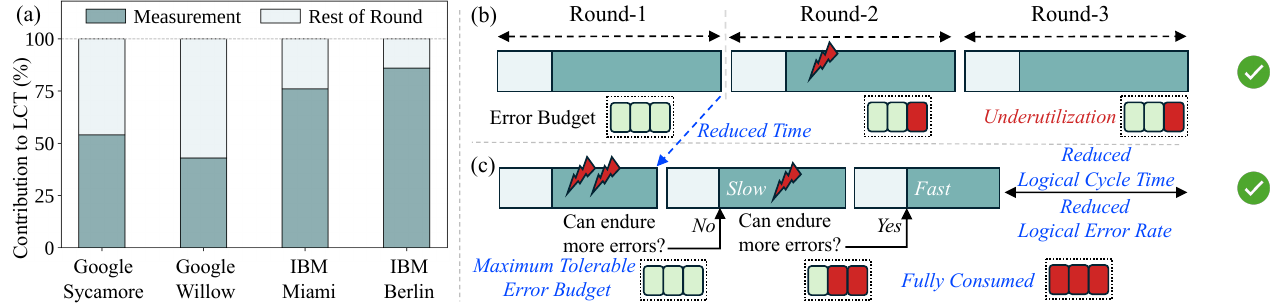}
  \vspace{-0.15in}
  \caption{(a)~Measurements account for 43--86\% of logical cycle times on superconducting platforms.
  (b)~In the baseline and \Dynamite{} Opt-1 Only, each syndrome extraction round uses the same readout latency, often leaving the error budget (number of tolerable errors) underutilized. (c)~\Dynamite{} Opt-1+Opt-2  dynamically selects fast or slow measurements per round based on the real-time consumption of the budget, reducing logical cycle time and error rates (\cmark).}
  \label{fig:intro}
\end{figure*}

Qubit measurement latencies dictate LCTs. Measurements are extremely slow and take one to two orders of magnitude longer than single and two-qubit gate operations.
For example, measurements account for 54\% of the LCT on Google Sycamore~\cite{google2023suppressing} and 43\% on Google Willow~\cite{google2025quantum_supp}. Similarly, on IBM Nighthawk, measurements take 2400\,ns, compared to gates that require only 32--161\,ns, and account for 76--86\% of the LCTs~\cite{ibm_nighthawk_calibration}. As shown in 
Figure~\ref{fig:intro}(a), measurements dominate LCTs across these state-of-the-art systems. \textit{Thus, reducing measurement latency is critical to lowering LCTs.}

Unfortunately, reducing measurement latencies at the device level is non-trivial because it directly impacts the fidelity of these operations. To measure a qubit in superconducting systems, a readout signal is sent to the qubit and the reflected signal trace from the measurement resonator is captured over a time period~\cite{krantz2019quantum}. Depending on the trace, a qubit is classified as a 0 or 1.
Shorter measurements are noisier, degrading the ability to correctly classify quantum states, whereas longer measurements increase the vulnerability of the qubit to idling errors and decoherence.
So, hardware providers calibrate these operations such that the latency achieves a sweet spot between the two types of errors and the operational fidelity of each qubit readout is maximized~\cite{bengtsson2024model, walter2017rapid, google2025quantum_supp}. Fault-tolerant programs utilize these calibrated per-qubit readout latencies to schedule \textit{all} measurement operations, as shown in Figure~\ref{fig:intro}(b).
However, calibrated latencies are often too conservative and sub-optimal for logical performance. 
This is because a logical cycle spans $d$ QEC rounds, each comprising single and two-qubit gates and parity qubit measurements, and therefore is subjected to the accumulated errors from all these operations. Moreover, data qubits are not measured and accumulate errors across all $d$ rounds before a correction is applied. 
As logical cycles span much longer than individual measurements, the likelihood of encountering decoherence increases with LCT, degrading LER.\textit{ Ideally, lower LCT and LER are desirable.}
Industry systems are already trading readout fidelity for faster measurements.
For example, IBM provides a mid-circuit measurement (\texttt{measure\_2}) that is ${\sim}65\%$ shorter but less accurate than its standard measurement on its heavy-hex processors~\cite{ibm_utility_scale_dynamic_circuits}, and a recent study shows that distance-5 compass codes on IBM Heron benefit from shorter measurements~\cite{lee2026scalable}.
\textit{This, faster but noisier measurements are becoming practical, which makes \Dynamite{} timely.}

%\textit{Ideally, we want to reduce both the logical cycle times (LCTs) and logical error rates (LERs), preserving the surface code's exponential LER suppression with increasing code distance $d$}.

%\newpage
We propose \textit{\Dynamite{}}, which addresses this challenge using two key insights. \textit{First, rather than using calibrated latencies that optimize per-qubit readout fidelity, a logical cycle should use a measurement latency fast enough to reduce both LCT and LER}. This is because 
although faster measurements may increase the error rate of individual readout operations, the collective error rate of all operations in the logical cycle and the likelihood of decoherence is still substantially reduced. Our experiments show that the LER is minimum at a measurement latency below the calibrated latency (baseline). For instance, lowering the measurement latency from 500\,ns to 200\,ns reduces the LER by 0.45$\times$, offering a 1.48$\times$ speedup for a distance-7 surface code on Google Sycamore. \Dynamite{} uses offline profiling to find this latency per device and distance.

\textit{Second, not all logical cycles are equal; some encounter more errors than others.} We study Hamming weight (number of 1s) distribution of the syndrome vectors (collective syndrome from $d$ rounds), which approximates the number of error events in a logical cycle. Logical cycles with HW below the \textit{Expected Hamming Weight (EHW)} barely affect the LER because decoders can accurately correct these events, and the EHW marks how much error a typical cycle contributes. For instance, on Google Sycamore at distance-7, 51\% of logical cycles finish below the EHW, and cycles above it are $6.5\times$ more likely to fail.
Given that faster measurements are acceptable provided the LER is not compromised, the code can tolerate HW up to a \textit{maximum tolerable error budget}, the largest end-of-logical-cycle syndrome vector HW below which the code can tolerate the errors.
Most logical cycles, however, operate far below this budget; this underutilized error budget can be used to lower logical cycle times through faster measurements.
However, this is non-trivial because existing schedulers select a uniform latency for all rounds at the start of the program, as shown in Figure~\ref{fig:intro}(b), whereas the errors encountered within a logical cycle are only known at runtime, so naively reducing the latency uniformly risks exceeding the error budget and degrading LER. Instead, \Dynamite{} leverages the fact that a logical cycle comprises $d$ rounds: it observes the cumulative HW after each round and adjusts subsequent latencies based on the error budget consumed and the number of rounds remaining. When the budget consumption is low, \Dynamite{} \textit{accelerates} subsequent rounds; when the budget is largely consumed, it \textit{protects} upcoming rounds by opting for slower measurements, as shown in Figure~\ref{fig:intro}(c).

Our experiments show that the lightweight \Dynamite{} reduces, on average, the LCT by 1.69$\times$ (up to 2.29$\times$) and the LER by 0.46$\times$ (down to 0.13$\times$), preserving the surface code's exponential error suppression with code distance.

Overall, this paper makes the following contributions.  
\begin{enumerate}[leftmargin=0cm,itemindent=.5cm,labelwidth=\itemindent,labelsep=0cm,align=left, itemsep=3pt, listparindent=0cm]

\item{We show that slow qubit measurements dictate LCTs and pose a critical bottleneck in reducing overall execution times.}

\item We show that calibrated latencies are \emph{conservative} because they optimize per-qubit readout fidelity, not logical performance: a lower latency reduces both LER and LCT.

\item We observe that each QEC code has an underutilized \textit{maximum tolerable error budget}, and propose a \textit{fine-grained measurement latency selection policy} to accelerate or protect each round in real-time within this budget, achieving additional speedup with a nominal impact on the LER.

\item We show that \Dynamite{} can also reduce \textit{physical overhead}: by lowering the LER, it can reach a target LER at a smaller code distance, using fewer physical qubits.

\item We demonstrate that \Dynamite{} is a lightweight design that accounts for less than \review{$0.2\%$}~logic utilization on off-the-shelf FPGAs and determines the latency for the upcoming round within \review{two clock cycles,}~facilitating real-time decisions.

\end{enumerate}

%\clearpage
\section{Background and Motivation}
\label{sec:background}

\subsection{Surface Code and Syndrome Extraction}
\label{bg:surface_code}

\begin{figure}[htp]
\centering
\includegraphics[width=\columnwidth]{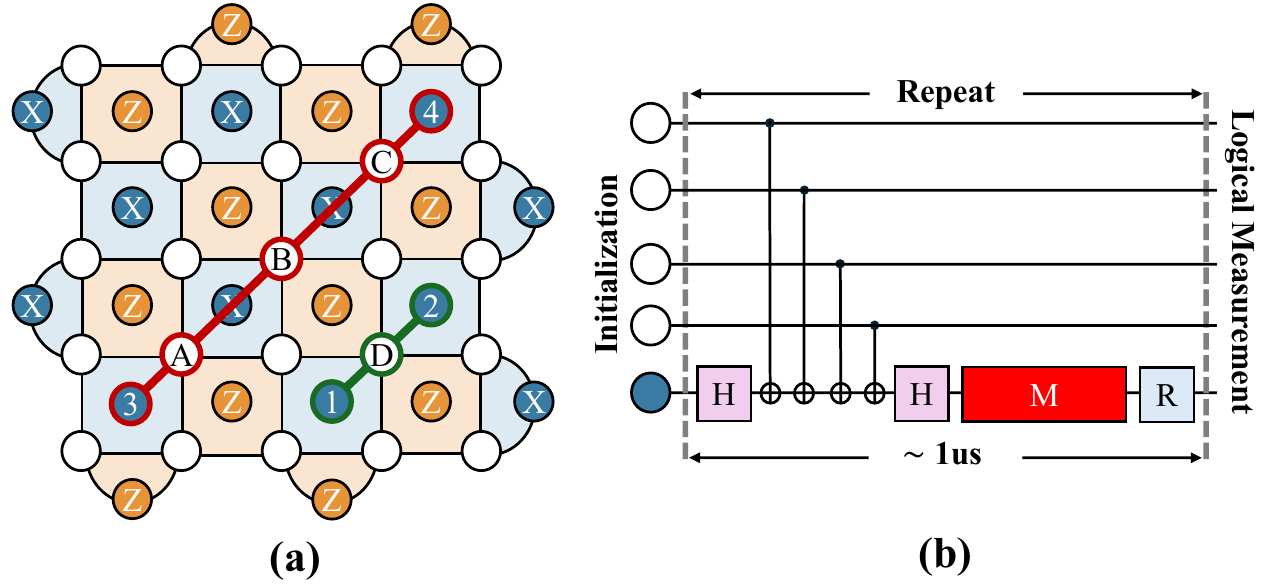}
\vspace{-0.25in}
\caption{(a)~Distance-\review{5}~surface code. $Z$ (orange) and $X$ (blue) stabilizers detect $X$ and $Z$ errors respectively. 
(b)~$X$ syndrome extraction circuit takes ${\sim}$1\,\textmu s to run on Google's systems, with latency dominated by measurement.}
\label{fig:surface_code}
\end{figure}

A distance-$d$ rotated surface code encodes a logical qubit into $d^2$ data and $d^2{-}1$ parity qubits in a lattice, as shown in Figure~\ref{fig:surface_code}(a). The parity qubits periodically measure multi-qubit stabilizers to detect errors: $X$ and $Z$ stabilizers detect phase ($Z$) and bit-flip ($X$) errors, respectively. \review{A data qubit error flips its adjacent parity qubits (qubits \circled{1} and \circled{2} flip due to error on qubit D). Errors on adjacent qubits result in longer error chains (qubits \circled{3} and \circled{4} flip due to errors on qubits A, B, and C). But even though more errors occur, only the endpoints of a chain flip. Parity qubits may also flip due to gate and measurement errors during this syndrome extraction process.}~
The parity bitstring, or syndrome, is used by a decoder to identify errors~\cite{dennis2001_surface_codes}; a distance-$d$ code corrects error chains of up to length $\lfloor(d-1)/2\rfloor$. Figure~\ref{fig:surface_code}(b) shows a syndrome extraction circuit. To tolerate errors in syndrome extraction, decoders use $d$ consecutive rounds, which form a \textit{syndrome vector} and constitute a \textit{logical cycle}.
%These rounds repeat from initialization until the data qubits are read out, the logical measurement; thus, data qubits remain unmeasured and hold the logical state throughout.
\review{X and Z errors are decoded independently using separate decoding graphs. The total syndrome Hamming weight of each graph depends on the number of error chains and not on the length of each chain. This is because irrespective of the chain length, only its end-points flip, resulting in a Hamming weight of 2 per chain.}~ 
On Google's hardware, a round takes ${\sim}1$ \textmu s~\cite{google2023suppressing, google_willow_spec, google2025quantum_supp}, giving a logical cycle time of ${\sim}d$ \textmu s; IBM's systems require even longer rounds due to slower measurements~\cite{ibm_nighthawk_calibration}.

% ------ FIGURE PLACEHOLDER ------
% --------------------------------

\subsection{Qubit Measurement Operations}
\label{bg:measurement_impact}
Measurements are critical to extract parity information and logical outcomes, but are slow and account for  
43--86\% of each syndrome extraction round on state-of-the-art superconducting systems~\cite{google2023suppressing, google2025quantum_supp, ibm_nighthawk_calibration}. Slow measurements limit us from reducing the runtime of fault-tolerant programs for two reasons. \textit{First}, programs rely on large distances~\cite{fowler2012surface, google2021exponential, google2023suppressing, google2025quantum} and logical cycle times increase linearly with distance. 
\textit{Second}, programs comprise millions of logical operations and only one logical operation can be executed per logical cycle~\cite{gidney2021_rsa_factorization, reiher2017elucidating, litinski2019game}. 
While fast measurements are desirable, they also have higher error-rates. For example, reducing measurement latency by $10\times$ increases its error rate by $123\times$ on Google Sycamore. Superconducting qubits use \textit{dispersive readout} in which a microwave signal is sent to the qubit and the resonance frequency of the measurement resonator connected to it experiences a shift based on its state. 
The qubit state is classified as a 0 or 1 by measuring this shift. This requires integrating the weak reflected signal against a background of thermal and amplifier noise~\cite{krantz2019quantum, bengtsson2024model}. 
Reducing latency degrades fidelity by overlapping the $0$/$1$ signal distributions.
This \emph{separation error}, $E_{\text{sep}}(t)$ is modeled using complementary error function (erfc) \cite{krantz2019quantum}, shown in Equation~\eqref{eq:separation_error}, where $a$ is the signal-to-noise ratio (SNR) growth rate and $t$ denotes latency.

% Google Sycamore
% measurement latency 500 -> error rate 0.1%
% measurement latency 50  -> error rate 12.2947%

\begin{equation}
    \label{eq:separation_error}
    E_{\text{sep}}(t) = \frac{1}{2} \text{erfc}\left( \frac{\sqrt{a \cdot t}}{2} \right)
\end{equation}

Conversely, slower measurements increase qubit relaxation via $T_1$ decay~\cite{krantz2019quantum, bengtsson2024model}. This \emph{relaxation error} $E_{\text{relax}}(t)$ follows Equation~\eqref{eq:relaxation_error}, with relaxation time $T_1$ and coefficient $c$.

\vspace{-0.1in}
\begin{equation}
    \label{eq:relaxation_error}
    E_{\text{relax}}(t) = c \cdot \left(1 - e^{-t / T_1}\right)
\end{equation}

\begin{figure}[htp]
\centering
\includegraphics[width=\columnwidth,]{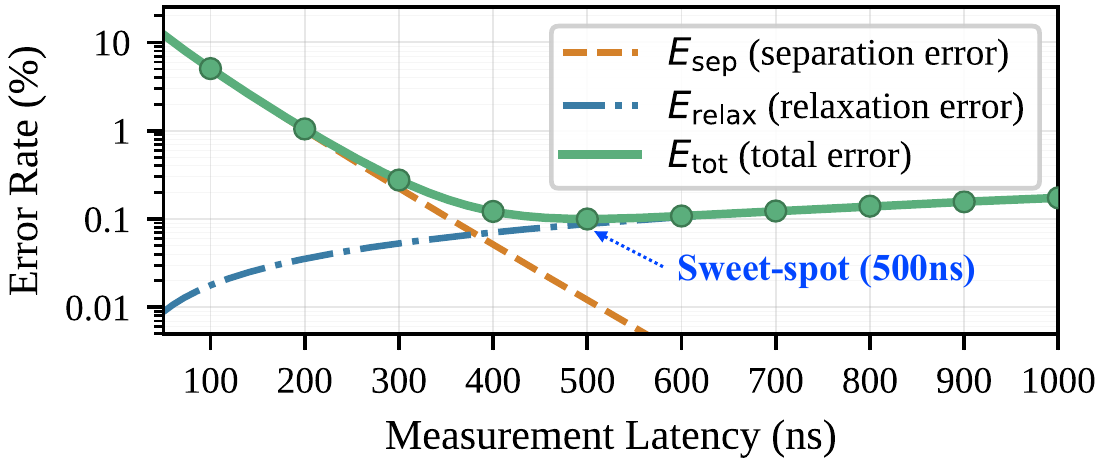}
\vspace{-0.30in}
\caption{Measurement latency is calibrated to minimize both separation and relaxation error and maximize readout fidelity (500\,ns on Google Sycamore).}
\label{fig:esep_erelax}
\vspace{-0.15in}
\end{figure}

Figure~\ref{fig:esep_erelax} shows the error rate with increasing measurement latencies. On Sycamore~\cite{google2023suppressing}, a latency of 500\,ns minimizes the overall error. Device providers calibrate the latency such that the readout fidelity of each qubit is maximized~\cite{bengtsson2024model, walter2017rapid, google2025quantum_supp}.

Qubit readout also faces errors from other sources, including low-temperature electronic devices and specialized instruments operating across thermal domains, sensitivity of the frequency shifts to environmental noise that lowers signal-to-noise ratios, and crosstalk from residual qubit-qubit couplings and frequency collisions~\cite{khezri2015qubit}. Thus, fast and accurate measurement at scale remains an open challenge that cannot be fully addressed at the device level. Similar trade-offs arise in other qubit technologies too, like neutral atoms and trapped ions. Nonetheless, using only separation and relaxation error offers an accurate first-order measurement error modeling~\cite{krantz2019quantum, bengtsson2024model}.

% ------ FIGURE PLACEHOLDER ------

%\subsection{Impact of $T_1/T_2$ Decoherence and Dynamical Decoupling}
\subsection{Measurement Latency Impacts Logical Performance}
\label{bg:decoherence_dd}

\begin{figure*}[!t]
\centering
\includegraphics[width=\textwidth]{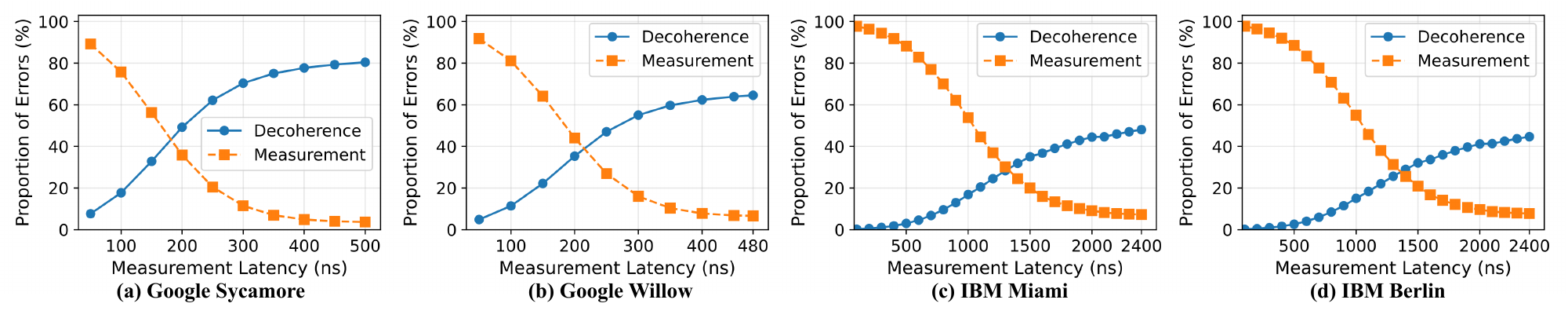}
\vspace{-0.25in}
\caption{Proportion of logical-cycle errors from decoherence and measurement versus measurement latency (distance-7, $p{=}0.1\%$, 100 million logical cycles). Measurement error dominates at short measurement latencies, whereas decoherence dominates at long measurement latencies.}
\label{fig:d7_100m_proportion_combined}
\end{figure*}

\begin{figure*}[!htp]
\centering
\includegraphics[width=\textwidth]{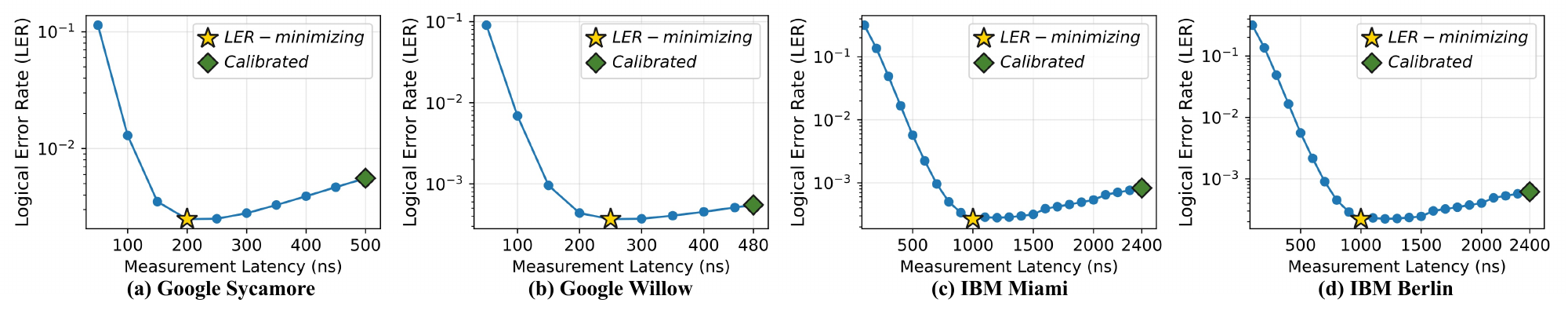}
\vspace{-0.25in}
\caption{Impact of measurement latency on logical error rate (LER) for distance-7 surface code on Google Sycamore, Google Willow, IBM Miami, and IBM Berlin ($p{=}0.1\%$, 100 million logical cycles). The LER is minimized at a much lower measurement latency than the calibrated latency.}
\label{fig:d7_100m_ler_combined}
\end{figure*}

Logical performance depends on logical cycle time (LCT) and logical error rate (LER), which dictate program latency and error resilience, respectively.
\textit{Ideally, lower LCT and LER are desirable.} Measurement latency determines the length of each QEC round and LCT. It also dictates the LER, which depends on the errors accumulated on both parity and data qubits operation errors during syndrome extraction and errors due to decoherence. 
A logical cycle spans $d$ QEC rounds, during which data qubits remain idle while parity qubits undergo measurement and reset, as shown in Figure~\ref{fig:surface_code}. 
This causes the data qubits to gradually lose their state through interactions with the environment or \textit{decoherence}. %Each qubit relaxes toward its ground state over a relaxation time $T_1$, and loses phase information over a dephasing time $T_2$~\cite{martinez2022decoherence}. 
The likelihood of decoherence increases exponentially with the idle time of qubits, which depends on the measurement latency. Consequently, longer measurements eventually degrade LER. Note that although idle errors can be reduced by applying dynamical decoupling 
pulses that keep the data qubits busy using low-cost single-qubit gates~\cite{google2023suppressing, google2025quantum}, decoherence cannot be fully eliminated.
Figure~\ref{fig:d7_100m_proportion_combined} shows how the dominant error source shifts from measurement to decoherence with increasing readout latency across four systems. When measurement latency is low, these operations account for most of the errors. In contrast, as the measurement latency increases, data qubit decoherence dominates the failures at the logical level.  
As a result, as shown in Figure~\ref{fig:d7_100m_ler_combined}, the LER is minimized at a measurement latency that is often lower than the calibrated latency, which only maximizes the fidelity of individual measurements. 

\subsection{Problem: Measurement Scheduling Is Sub-Optimal:}
Measurement scheduling in fault-tolerant programs suffers from two key drawbacks. \textit{First}, calibrated latencies do not optimize for logical performance. \textit{Second}, all $d$ QEC rounds of a logical cycle use this statically chosen latency uniformly. This limits us from improving LCT because our experiments show that not all logical cycles encounter the same level of errors; some are more erroneous than others. But even if a logical cycle encounters very few errors and could benefit from even lower readout latencies to reduce LCT because these cycles could tolerate slightly higher measurement error rates, static and uniform latency selection limits us from doing so. 

\bluebox{\textbf{\textit{Goal}} Our goal is to reduce both logical error rate and logical cycle time by selecting the most suitable measurement latency per logical qubit and cycle.}

\section{Dynamite: Key Insights}
\label{sec:insights}

\begin{figure*}[!htp]
  \centering
  \includegraphics[width=\textwidth]{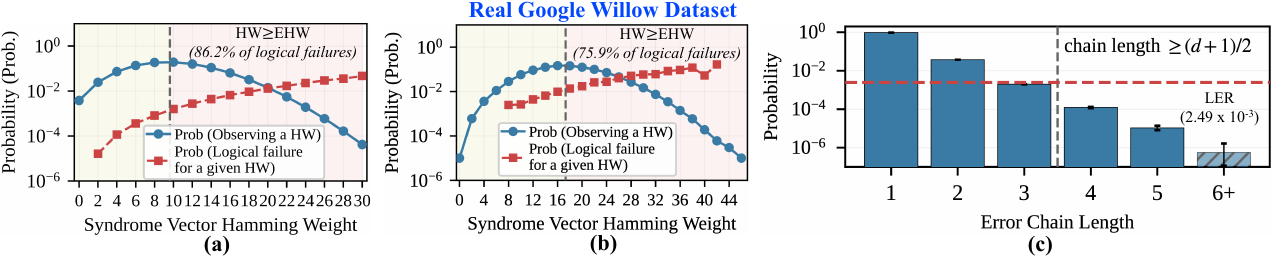}
  \vspace{-0.25in}
  \caption{\review{Distribution of syndrome vector HW and conditional logical failure rate for Google Sycamore at latency=200\,ns ($p$=0.1\%, 100M logical
  cycles, EHW=9.62) for distance 7 surface code (a) based on simulation and (b) from \textbf{real Google Willow dataset}~\cite{google2025quantum, google2025quantum_supp} (100,000 logical cycles of 10 rounds each, EHW=17.3). (c) Distribution of error chain lengths for Google Sycamore at latency=200\,ns ($p$=0.1\%, 1M logical cycles). Low HW cycles are common yet rarely fail, whereas high HW cycles are less likely yet fail more often; vertical line marks $(d{+}1)/2$, the shortest length that can cause a logical failure.}}
  \label{fig:hw_dist_concept_error_chain}
  \vspace{-0.1in}
\end{figure*}

We propose \textit{\Dynamite{}}, a lightweight real-time measurement latency selection policy to improve logical performance. We discuss the key insights of \Dynamite{} in this section. 

\subsection{Select Readout Latency To Optimize Logical Performance}
\label{insight:tradeoff}

Although faster measurements may make individual qubit readouts slightly more error-prone, they are often optimal for logical performance, as shown in Figure~\ref{fig:d7_100m_ler_combined}. For example, the LER of a distance-7 surface code on Google Sycamore is reduced by $0.45\times$ when we use a measurement latency of 200\,ns as opposed to the calibrated 500\,ns. This also reduces syndrome extraction time from 921\,ns to 621\,ns, which decreases the LCT from 6.4\,$\mu$s to 4.3\,$\mu$s, yielding a $1.48\times$ speedup. %, as the LER falls to $0.45\times$ the baseline, from $5.57{\times}10^{-3}$ to $2.49{\times}10^{-3}$.
We observe this trend across four different quantum platforms and every code distance. Thus, rather than naively relying on the calibrated latencies provided by the hardware vendor to schedule measurements, FTQCs must identify the latency specific to each code that reduces both LER and LCT.

\bluebox{\textbf{\textit{Insight-1:}} Logical cycles should use a measurement latency fast enough to reduce both LCT and LER rather than optimizing per-qubit readout fidelity.}

%%% section 3.2 %%%%%%%
\subsection{Select Measurement Latency Dynamically And At A Finer Granularity To Further Improve Logical Performance}
\label{insight:budget}

%Not all logical cycles are equal; some experience more errors than others. 
An error chain flips at most two parity qubits irrespective of length. 
The \textit{Hamming Weight (HW)} of a syndrome vector is the number of 1s in its bitstring and approximates the number of error events in a logical cycle. The \textit{Expected Hamming Weight (EHW)} is the average number of bit flips in a syndrome vector, which depends on errors encountered and code distance.
We measure the \textit{conditional logical failure rate} (P$_\text{logical\_fail}$) which is the probability that a logical cycle fails given a syndrome vector HW.
Figure~\ref{fig:hw_dist_concept_error_chain}(a) shows the HW and conditional logical failure rate over 100M logical cycles for a distance-7 surface code on Sycamore when LER-minimizing latency of 200\,ns is used for measurements. The likelihood of observing a HW is non-uniform and probability of logical failure increases with HW, with 86.2\% of failures arising from cycles with HW above EHW. Table~\ref{tab:ehw_budget} shows the EHW, LER, and P$_\text{logical\_fail}$.

\vspace{-0.1in}
\begin{table}[htp]
      \centering
      \caption{EHW, LER, and P$_\text{logical\_fail}$ for Google Sycamore at latency minimizing LER ($p{=}0.1\%$, 100 million logical cycles).}
      \label{tab:ehw_budget}
      \footnotesize
      \setlength{\tabcolsep}{3pt}\renewcommand{\arraystretch}{1.2}
      \begin{tabular}{@{}|c|c|c|c|c|c|@{}}
      \hline
      \makecell{$d$} & \makecell{EHW} & \makecell{LER}
       & \makecell{Prob.\\HW$\leq$EHW}
       & \makecell{P$_\text{logical\_fail}$\\HW$\leq$EHW}
       & \makecell{P$_\text{logical\_fail}$\\HW$>$EHW} \\
      \hline\hline
       3 &  0.56 & 1.12$\times10^{-2}$ & 67.31\% & 2.86$\times10^{-5}$ & 3.43$\times10^{-2}$ \\ \hline
       5 &  3.01 & 5.86$\times10^{-3}$ & 60.94\% & 1.35$\times10^{-3}$ & 1.29$\times10^{-2}$ \\ \hline
       7 &  9.62 & 2.49$\times10^{-3}$ & 50.96\% & 6.73$\times10^{-4}$ & 4.37$\times10^{-3}$ \\ \hline
       9 & 20.79 & 9.83$\times10^{-4}$ & 49.96\% & 3.20$\times10^{-4}$ & 1.65$\times10^{-3}$ \\ \hline
      11 & 38.32 & 3.79$\times10^{-4}$ & 52.17\% &         
       1.46$\times10^{-4}$& 6.33$\times10^{-4}$\\ \hline
      13 & 63.65 & 1.41$\times10^{-4}$ & 50.42\% & 
       5.79$\times10^{-5}$& 2.26$\times10^{-4}$\\ \hline
      \end{tabular}
\end{table}

On Sycamore at distance-7, using 200\,ns measurement latency with an EHW of $9.62$, logical cycles with HW up to EHW ($51\%$ of total) fail with a probability of $6.73{\times}10^{-4}$, $3.7\times$ lower than LER of $2.49{\times}10^{-3}$. In contrast, logical cycles with HW above EHW fail with probability $4.37{\times}10^{-3}$, $1.8\times$ higher than LER.
Thus, logical cycles with HW up to the EHW barely affect the LER because the decoder can correct the errors, whereas they become $6.5\times$ more likely to fail once their HW exceeds it. We observe similar trends across four different quantum computers. Logical cycles with HW above the EHW are $3.9\times$ to $5152\times$ more likely to fail than those with HW up to EHW and account for $79.3\%$ to $99.9\%$ of the logical failures.
Moreover, logical cycles with HW far below the EHW denote wasted error budget and underutilization of decoding capabilities. These low-HW logical cycles could employ even faster measurements, albeit at the expense of more error-prone readouts, because they can tolerate \textit{slightly} more errors. The impact on the LER would be negligible as long as the syndrome vector HW does not exceed the EHW.

\review{We make similar observations on real hardware too. Figure~\ref{fig:hw_dist_concept_error_chain}(b) shows syndrome vector HW distribution and conditional logical failure rates (P$_\text{logical\_fail}$) from Google Willow's publicly available dataset for a distance-7 surface code~\cite{google2025quantum, google2025quantum_supp}. Note that logical cycles in this dataset are much longer ($d{+}3$ rounds as opposed to conventional $d$ rounds) and subjected to much higher measurement error rates (0.8\% as opposed to 0.1\% used in our simulations to represent future devices). 53\% of logical cycles have HW up to the EHW, and logical cycles with HW above the EHW are 3.6$\times$ more likely to fail and account for 75.9\% of the logical failures. \textbf{\textit{Thus, our observation in simulation is consistent with real hardware. }}

\vspace{0.05in}
\noindent \textbf{Is Error Budget Indeed Present?} The code distance provisions for \textit{worst-case} errors. However, errors are stochastic and not all logical cycles encounter the same level of errors. \textit{For example, using distance 7 guarantees an ability to correct error-chains of length up to 3, \textbf{but it does not mean} that all logical cycles in practice will only and always encounter error chains of length 3.} More often, syndromes are sparse, with a few localized errors. Figure~\ref{fig:hw_dist_concept_error_chain}(c) shows the distribution of error chain lengths for a distance 7 code-- 96.0\%, 3.7\%, and 0.2\% of error chains correspond to lengths 1, 2, and 3, respectively. 
Many prior works~\cite{das2022lilliput, das2022afs, vittal2023astrea, delfosse2020hierarchical, alavisamani2024promatch, knapen2026pinball, ravi2023better}, including real hardware data have made similar observations~\cite{google2025quantum, google2025quantum_supp}. \textbf{\textit{Thus, there is indeed an error budget.}} But this budget is \textit{not uniformly available in all logical cycles}; only cycles encountering fewer errors than code capacity offer this budget. If we use the budget intelligently to shorten rounds, it improves both runtime and LER, making it possible to reduce code distance (further discussion in Section~\ref{sec:cross_device}).}

\bluebox{\textbf{\textit{Insight-2:}} Not all logical cycles are equal; cycles with HW below EHW are far less likely to fail than those above it. We can use faster measurements without impacting LER, provided the HW stays below EHW.}
\begin{figure*}[htp]
  \centering
  \includegraphics[width=\textwidth]{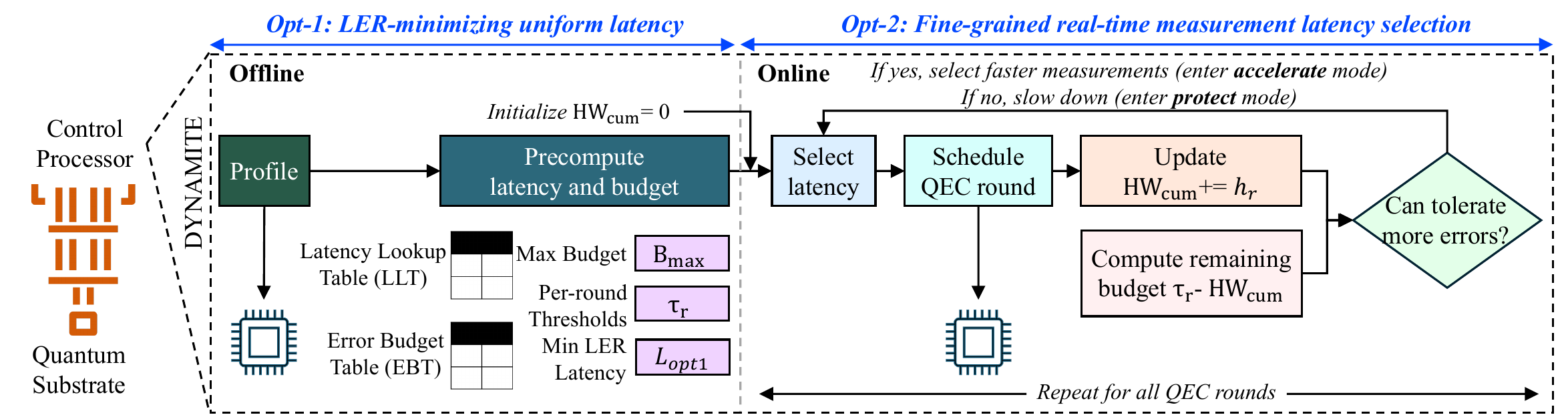}
  \caption{\Dynamite{} sits at the control processor and comprises two phases. The offline phase profiles the latency versus LER trade-off to find the LER-minimizing latency $L_\text{opt1}$, and profiles the conditional logical failure rate to compute the per-round error budget thresholds $\tau_r$ and the maximum tolerable error budget $B_\text{max}$ for each device and code distance. The online phase tracks the cumulative syndrome vector HW ($HW_{\text{cum}}$) at each round and, from the remaining budget ($\tau_r - HW_\text{cum}$), selects the fastest measurement latency for the next round.}
  \label{fig:dynamite_overview}
\end{figure*}
\section{Our Proposal: \Dynamite{}}
\label{sec:design}

We propose \textit{\Dynamite{}}, a fine-grained dynamic measurement latency selection policy for FTQCs. Figure~\ref{fig:dynamite_overview} gives an overview of \Dynamite{}. It comprises (1)~an offline phase that profiles the measurement latency versus fidelity trade-off and identifies the latency that minimizes the LER, maximum tolerable error budget, and largest tolerable error budget per round, and (2)~a lightweight online phase that selects the measurement latency for each round from the error budget consumed so far. Table~\ref{tab:parameters} lists \Dynamite{}'s parameters before we describe our design.

\begin{table}[htb]
        \begin{center}
        \caption{Parameters used in \Dynamite{}}
        \setlength{\tabcolsep}{1.0mm}
        \renewcommand{\arraystretch}{1.2}
        \label{tab:parameters}
        \begin{tabular}{ |c|c| }
        \hline
        Parameter & Description \\
        \hline\hline
        \multirow{2}{*}{$\text{EHW}(L)$} & Expected syndrome vector Hamming Weight over\\
        & a logical cycle at measurement latency $L$\\
        \hline
        $\mu_L$ & EHW per round $=\text{EHW}(L)/d$\\
        \hline
        $h_r$ & Observed syndrome vector HW of round $r$\\
        \hline
        $HW_{\text{cum}}$ & Cumulative syndrome vector HW\\
        \hline
        $\tau_r$ & Largest tolerable HW$_\text{cum}$ through round $r$\\
        \hline
        $B_{\text{max}}$ & Maximum tolerable error budget\\
        \hline
        $L_{\text{opt1}}$ & Measurement latency that minimizes the LER\\
        \hline
        $L_\text{base}$ & Baseline calibrated measurement latency\\
        \hline
        \end{tabular}
        \end{center}
        \vspace{-0.2in}
\end{table}

\subsection{Offline Profiling (Optimization-1 or Opt-1)}
\label{design:error_model}
%To assess the feasibility of using lower measurement latencies, the offline phase profiles (1)~the trade-off between measurement latency and fidelity, (2)~the error budget for each latency, and (3)~the latency that minimizes the LER and its maximum tolerable error budget.
The offline profiling phase comprises three tasks.

\subsubsection{\textbf{Measurement Latency and Fidelity Profiling}} \hfill

\Dynamite{} prepares a \textit{Latency Lookup Table (LLT)} that stores the measurement error rate for a range of latencies up to the calibrated measurement latency $L_\text{base}$. For our default implementation, we use 50\,ns to $L_\text{base}$ in 50\,ns steps for Google devices and from 100\,ns to $L_\text{base}$ in 100\,ns steps for IBM devices.
The entries are obtained by calibrating the qubits for each latency value. However, we lack the low-level system access to conduct these experiments. We therefore use an analytical model based on device-level studies~\cite{krantz2019quantum, bengtsson2024model} instead.
To compute the error rate at a specific latency, we model the measurement error $p_\text{meas}(t)$ as the sum of separation $E_\text{sep}(t)$ and relaxation $E_\text{relax}(t)$ errors, as shown in~Equation~\eqref{eq:meas_error}.

\begin{equation}
    \label{eq:meas_error}
        p_{\text{meas}}(t) = \underbrace{\tfrac{1}{2}\,
        \text{erfc}\!\left(\tfrac{\sqrt{a \cdot t}}{2}\right)}_{E_{\text{sep}}(t)}
        + \underbrace{c \cdot \left(1 - e^{-t / T_1}\right)}_{E_{\text{relax}}(t)}
\end{equation}

Recall that separation error depends on the SNR growth rate $a$, whereas relaxation error depends on the relaxation time $T_1$ and pre-factor $c$~\cite{bengtsson2024model}.
As device providers do not report parameters $a$ and $c$, we compute them using the condition that $p_\text{meas}(t)$ is minimized at $t_b{=L_\text{base}}$ for each system. % and (2)~$p_\text{meas}$ at 500\,ns is about 0.1\% (same as $p$). 
This requires the total error derivative to be 0, given by Equation~\eqref{eq:optimality}:

\begin{equation}
        \label{eq:optimality}
        \frac{d}{dt} p_\text{meas}(t) \Big|_{t=t_b}
        = -\frac{1}{4\sqrt{\pi}} \sqrt{\frac{a}{t_b}}\, e^{-at_b/4} + \frac{c}{T_1}\, e^{-t_b/T_1} = 0
\end{equation}

Solving this yields $a{=}0.0539\,/\text{ns}$ and $c{=}0.0356$ for Google Sycamore with $p_\text{meas}$ at 500\,ns as 0.1\%. We compute $a$ and $c$ similarly for other systems used in our paper: Google Willow, IBM Miami, and IBM Berlin~\cite{google2025quantum_supp, ibm_nighthawk_calibration}. We use these derived values to construct the LLT, as shown in Table~\ref{tab:error_rate_table}.

% --- Table: Measurement Error Rates ---
\begin{table}[!htp]
    \centering \small
    \setlength{\tabcolsep}{0pt}
    \renewcommand{\arraystretch}{1.0}
    \setcellgapes{0pt}\makegapedcells
    \caption{Measurement Latency Lookup Table (LLT): $p_\text{meas}$ versus latency for the four devices ($p{=}0.1\%$). Bold marks $p_\text{meas}$ at $L_\text{base}$; ``---'' marks latencies above $L_\text{base}$ or off the device grid.}
    \label{tab:error_rate_table}
    \begin{tabular}{|c|c|c||c|c|c|}
    \hline
     & \makecell{Google\\Sycamore} & \makecell{Google\\Willow}
     & & \makecell{IBM\\Miami} & \makecell{IBM\\Berlin} \\
    \hline\hline
    $a$ (/ns) & 0.0539 & 0.0561 & $a$ (/ns) & 0.0112 & 0.0112 \\ \hline
    $c$       & 0.0356 & 0.1248 & $c$       & 0.1106 & 0.1215 \\
    \hline\hline
    \makecell{Latency\\(ns)} & \multicolumn{2}{c||}{$p_\text{meas}$ (\%)}
     & \makecell{Latency\\(ns)} & \multicolumn{2}{c|}{$p_\text{meas}$ (\%)} \\
    \hline
     50 & 12.295 & 11.829         &  100 & 22.701 & 22.702 \\
    100 &  5.051 &  4.720         &  200 & 14.486 & 14.487 \\
    150 &  2.245 &  2.042         &  300 &  9.741 &  9.742 \\
    200 &  1.048 &  0.931         &  400 &  6.725 &  6.726 \\
    250 &  0.516 &  0.451         &  500 &  4.720 &  4.721 \\
    300 &  0.276 &  0.241         &  600 &  3.352 &  3.353 \\
    350 &  0.168 &  0.151         &  700 &  2.404 &  2.404 \\
    400 &  0.122 &  0.114         &  800 &  1.738 &  1.738 \\
    450 &  0.104 &  0.101         &  900 &  1.266 &  1.267 \\
    480 &  ---   & \textbf{0.100} & 1000 &  0.931 &  0.931 \\
    500 & \textbf{0.100} & ---    & 1100 &  0.690 &  0.691 \\
    --- & ---    & ---            & \dots & \dots & \dots \\
%   --- & ---    & ---            & 1200 &  0.518 &  0.518 \\
%   --- & ---    & ---            & 1300 &  0.394 &  0.394 \\
%   --- & ---    & ---            & 1400 &  0.305 &  0.305 \\
%   --- & ---    & ---            & 1500 &  0.241 &  0.241 \\
%   --- & ---    & ---            & 1600 &  0.196 &  0.196 \\
%   --- & ---    & ---            & 1700 &  0.163 &  0.163 \\
%   --- & ---    & ---            & 1800 &  0.140 &  0.140 \\
%   --- & ---    & ---            & 1900 &  0.124 &  0.124 \\
%   --- & ---    & ---            & 2000 &  0.114 &  0.114 \\
%   --- & ---    & ---            & 2100 &  0.107 &  0.107 \\
%   --- & ---    & ---            & 2200 &  0.103 &  0.103 \\
    --- & ---    & ---            & 2300 &  0.101 &  0.101 \\
    --- & ---    & ---            & 2400 & \textbf{0.100} & \textbf{0.100} \\
    \hline  
    \end{tabular}
\end{table}

\subsubsection{\textbf{Average Error Budget Profiling}} \hfill

Next, \Dynamite{} computes the \textit{average error budget}
per code distance by profiling the $\text{EHW}$ for each latency ($L$) from Monte Carlo simulations of 100 million logical cycles, to form the \textit{Error Budget Table (EBT)}. The average error budget approximates the average number of error events that will be observed while using $L$.
The $\text{EHW}$ is profiled for each quantum system. Table~\ref{tab:error_budget} shows the EBT of Google Sycamore as an example. 
%As expected, the EHW increases as the latency decreases because faster measurements are noisier.
%But near $L_\text{opt1}$, this increase is only modest. For instance, on Google Sycamore at $d{=}7$, reducing the measurement latency from $L_\text{opt1}$ to the next faster step raises the EHW by only $1.27\times$.
%This has a minimal impact on the LER: for logical cycles that stay below the budget, the extra HW falls within the range the decoder already tolerates.
\Dynamite{} also records the conditional failure rate (P$_\text{logical\_fail}$) versus the cumulative syndrome HW ($HW_\text{cum}$) and the LER for each latency in the EBT. This information is used to identify the most suitable latency for each upcoming round in the online phase, as will be explained later.
%the next step uses these to determine the maximum tolerable error budget and the latency $L_\text{opt1}$ that minimizes the LER.

\begin{table}[!htp]
        \centering\small
        \setlength{\tabcolsep}{2pt}\renewcommand{\arraystretch}{1.0}
        \caption{Error Budget Table (EBT): EHW($L$) versus measurement latency and code distance on Google Sycamore ($p{=}0.1\%$).}
        \label{tab:error_budget}
        \begin{tabular}{|c|c|c|c|c|c|c|}
        \hline
        $L$ (ns) & $d{=}3$ & $d{=}5$ & $d{=}7$ & $d{=}9$ & $d{=}11$ & $d{=}13$ \\
        \hline\hline
        250 & 0.56 &  3.01 &  8.69 & 18.96 & 35.15 & 58.63 \\
        200 & 0.66 &  3.38 &  9.62 & 20.79 & 38.32 & 63.65 \\
        150 & 0.89 &  4.40 & 12.22 & 26.05 & 47.60 & 78.56 \\
        100 & 1.46 &  6.86 & 18.63 & 39.17 & 70.91 & 116.27 \\
         50 & 2.78 & 12.73 & 34.11 & 71.15 & 128.10 & 209.21 \\
        \hline
        \end{tabular}
\end{table}

\subsubsection{\textbf{LER Optimal Latency and Tolerable Budget Profiling}} %\hfill
\label{sec:sdynamite}

Using the first insight, \Dynamite{} identifies the measurement latency ($L_\text{opt1}$) that minimizes the LER for each machine and distance $d$, as explained by Equation~\eqref{eq:staticsel}. 
\Dynamite{} uses this latency as the default latency to schedule each measurement. 

%For each device and code , \Dynamite{} profiles the LER across LLT latencies.
%As latency drops below the baseline $L_\text{base}$, measurement error $p_\text{meas}$ rises while shorter idling lowers decoherence. This reduction initially outweighs the higher measurement error, keeping LER at or below the baseline; once $p_\text{meas}$ dominates at shorter latencies, the LER increases, creating an intermediate minimum.
% defines $L_\text{opt1}$ as the latency that minimizes the LER at distance-$d$.
%with the resulting values summarized in Table~\ref{tab:L_opt1}.

\begin{equation}
        \label{eq:staticsel}
        L_\text{opt1} = \operatorname{arg\,min}_{L}\ \text{LER}(d, L)
\end{equation}

\begin{figure}[htp]
\centering
\includegraphics[width=\columnwidth]{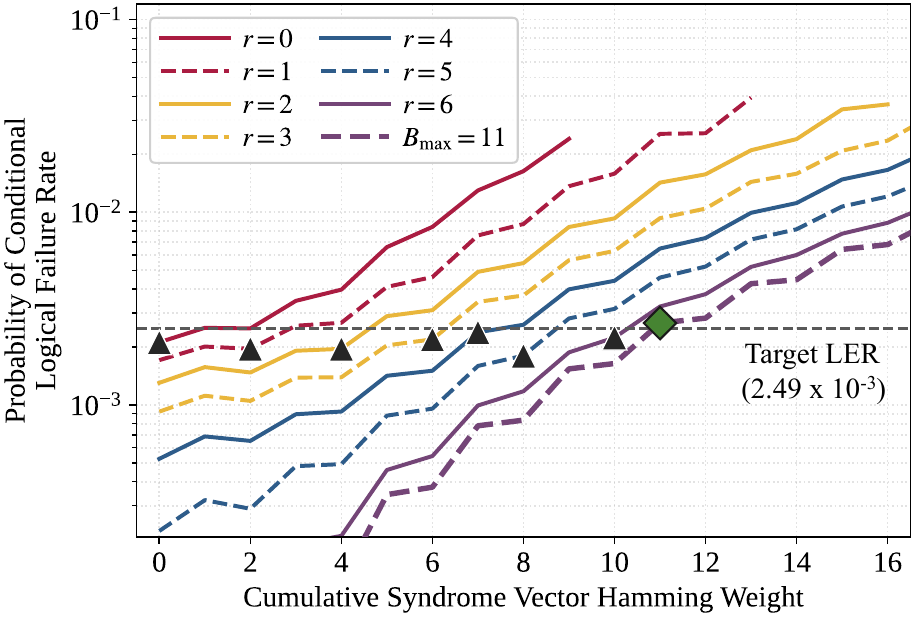}
\vspace{-0.25in}
\caption{Per-round conditional logical failure rate versus cumulative syndrome vector Hamming weight (Google Sycamore, distance-7, $L_\text{opt1}{=}200$\,ns, $p{=}0.1\%$, 100 million logical cycles).}
%Each round-$r$ curve and the logical cycle curve cross the LER obtained using $L_\text{opt1}$ (dashed); these crossings define the maximum tolerable error budget for each round ($\triangle$) and for logical cycle ($\diamond$).}
\label{fig:tau_per_round}
\end{figure}

%This single latency improves both the LER and the speedup over the baseline. We evaluate this uniform latency policy as \Dynamite{} (Opt-1 Only).
\Dynamite{} then profiles the maximum tolerable error budget threshold for each round ($\tau_r$) and the logical cycle ($B_\text{max}$). To compute this, it profiles the HW distribution at the end of each round and for the entire logical cycle. The error budget threshold for the $r^{\text{th}}$ round, where $0\leq r\leq d-1$, is the largest syndrome vector HW up to that round, such that the conditional failure rate ($P_{\text{logical\_fail}}$) remains below the LER obtained using $L_\text{opt1}$. 
Similarly, $B_\text{max}$ is the maximum syndrome vector HW tolerable at the end of the logical cycle such that $P_{\text{logical\_fail}}$ remains below the LER obtained at $L_\text{opt1}$.
%Figure~\ref{fig:tau_per_round} illustrates this process for Google Sycmaore at distance-7, and Table~\ref{tab:tau} shows the $L_\text{opt1}$, $B_\text{max}$, and $\tau_r$ for Google Sycamore at each distance-$d$. The EBT stores these profiling results.
Figure~\ref{fig:tau_per_round} illustrates this process for Google Sycamore at distance-7, and Table~\ref{tab:tau} shows the profiled $L_\text{opt1}$, $B_\text{max}$, and $\tau_r$ for Google Sycamore stored in the EBT at each distance-$d$.

\begin{table}[!htp]
%\vspace{-0.15in}
      \centering
      \caption{$L_\text{opt1}$, $B_\text{max}$, and $\tau_r$ entries in EBT for various code distances on Google Sycamore ($p{=}0.1\%$, 100 million logical cycles).}
      \label{tab:tau}
      \footnotesize
      \setlength{\tabcolsep}{3pt}\renewcommand{\arraystretch}{1.15}
      \begin{tabular}{|c|c|c|l|}
      \hline
      % Device & 
      $d$ & $L_\text{opt1}$ (ns) & $B_\text{max}$ & $\tau_0,\dots,\tau_{d-1}$ \\
      \hline\hline
      % \multirow{6}{*}{\makecell{Google\\Sycamore}} & 
 3 & 250 &  1 & 0, \phantom{0}0, \phantom{0}0 \\
 % & 
 5 & 250 &  5 & 0, \phantom{0}2, \phantom{0}2, \phantom{0}2, \phantom{0}3 \\
 % & 
 7 & 200 & 11 & 0, \phantom{0}2, \phantom{0}4, \phantom{0}6, \phantom{0}7, \phantom{0}8, 10  \\
 % & 
 9 & 200 & 23 & 2, \phantom{0}4, \phantom{0}7, 10, 12, 14, 16, 19, 21  \\
 % & 
11 & 200 & 42 & 3, \phantom{0}8, 10, 14, 18, 21, 25, 28, 32, 35, 38 \\
 % & 
13 & 200 & 67 & 5, 10, 16, 20, 25, 29, 34, 40, 44, 49, 55, 58, 64  \\
      \hline
      \end{tabular}\vspace{-0.1in}
\end{table}

% ------ FIGURE: d=7 HW distribution ------

\subsection{Online Latency Selection (Optimization-2 or Opt-2)}
\label{sec:adaptive}

In online phase, \Dynamite{} exploits the second insight and selects the measurement latency for each QEC round in real time based on the consumption of the maximum tolerable error budget ($B_\text{max}$) up until that round and the number of remaining rounds. Recall that in both the baseline and \Dynamite{} using only Opt-1 applies the calibrated measurement latency and $L_\text{opt1}$ respectively uniformly to all measurements. This coarse-granularity and static approach under-utilizes the available error budget, as our experiments show that several logical cycles end up with syndrome HWs far below budget. Intuitively, these logical cycles could have used even measurement latencies below $L_\text{opt1}$. But, reducing measurement latency is challenging; the speedup achievable would be lower if the reduction is low, whereas measurement errors would dominate and increase LER if the reduction is too high. So, faster measurements must be judiciously used for only those logical cycles (with low syndrome HW) that would benefit. The latency should also be chosen such that the logical cycle would still be able to tolerate the increased measurement error rate. Other logical cycles that are already experiencing moderately high or high HW should continue to use $L_\text{opt1}$ to ensure that the error budget is not exceeded and LER is not compromised. This is non-trivial with static and uniform latency selection because error information is only available at runtime. 

\Dynamite{} addresses these challenges using fine-grained dynamic measurement latency selection for each QEC round depending on the consumption of the error budget in real-time. The number of errors encountered varies from round to round: some rounds add little to cumulative HW ($HW_\text{cum}$) and leave budget to spare, whereas others consume more. \Dynamite{}'s fine-grained latency selection adapts the latency for each round based on the history of errors observed within the current logical cycle to exploit this unused budget. This tracking also ensures that the budget is not exceeded.  

\Dynamite{} starts each logical cycle by scheduling measurements at $L_\text{opt1}$ latency and keeps $HW_\text{cum}$ within the budget threshold ($\tau_r$) at the end of each round, so that $HW_\text{cum}$ reaches close to $B_\text{max}$ by the end of the logical cycle.
%It schedules the $d$ syndrome extraction rounds under $B_\text{max}$.
The EHW per round at latency $L$ is $\mu_L{=}\text{EHW}(L)/d$. \Dynamite{} monitors the errors within a logical cycle using the $HW_\text{cum}$ counter, initialized to $0$ at the start of each cycle. To schedule round $r$, \Dynamite{} consults $\tau_r$ and picks the fastest latency whose $\mu_L$ keeps $HW_\text{cum}$ within $\tau_r$, which is the fastest $L{\le}L_\text{opt1}$ satisfying Equation~\eqref{eq:find_latency}. Restricting the search to $L{\le}L_\text{opt1}$ lets \Dynamite{} accelerate rounds with spare error budget for speedup, whereas $\tau_r$ ensures that LER does not degrade.
\begin{equation}
        \label{eq:find_latency}
        \mu_L = \frac{\text{EHW}(L)}{d} \;\le\; \tau_r - HW_\text{cum},
        \qquad L \le L_\text{opt1}
\end{equation}

Once round $r$ executes, \Dynamite{} operates in two modes, \textit{accelerate} and \textit{protect}, to dynamically adjust measurement latency based on the real-time error budget ($\tau_r - HW_\text{cum}$). \Dynamite{} progressively pushes for faster measurements when the budget consumption is low, which corresponds to the \textit{accelerate} mode. In contrast, as $HW_\text{cum}$ approaches $\tau_r$ and the remaining budget shrinks, \Dynamite{} opts for slower measurements. When no latency fits the remaining budget, \Dynamite{} enters \textit{protect} mode and reverts to $L_\text{opt1}$ for the next round.

We explain this with a distance-7 surface code on Google Sycamore whose $L_\text{opt1}{=}200$\,ns and $B_\text{max}{=}11$. Assume that while scheduling round 4, $HW_\text{cum}$ is 4; the headroom to the safe threshold is then $\tau_4{-}HW_\text{cum}{=}7{-}4{=}3$.
This permits \Dynamite{} to use 100\,ns for round 4, because its expected per-round HW $\text{EHW}(100)/7{=}2.66$ stays below 3.
Now assume that round 4 is a noisy round with 4 errors, raising $HW_\text{cum}$ to 8. When scheduling round 5, the headroom $\tau_5{-}HW_\text{cum}{=}8{-}8{=}0$ becomes $\le0$, so \Dynamite{} enters protect mode, forcing $L_\text{opt1}$ for round 5. If round 5 then observes zero errors, the headroom $\tau_6{-}HW_\text{cum}{=}10{-}8{=}2$ grows again and \Dynamite{} enters accelerate mode in round 6, using faster latency 150\,ns.

We use a Minimum Weight Perfect Matching (MWPM) decoder where \textit{weights} correspond to the probability of each error event.
\Dynamite{} varies the latency across rounds and notifies the decoder of each round's latency, so the decoder can use that round's error probabilities for accurate decoding.
%Algorithm~\ref{alg:dynamite_online} at the end of the paper summarizes \Dynamite{}.
%\newpage
\section{Evaluation Methodology}
\label{sec:methodology}
In this section, we discuss our evaluation methodology. 

\subsection{Device Parameters}
\label{methodology:device_parameters}

We use four state-of-the-art superconducting systems for our evaluations: Google Sycamore (72-qubit)~\cite{google2023suppressing}, Google Willow (105-qubit)~\cite{google2025quantum_supp}, and two 120-qubit IBM Quantum Nighthawk r1 systems, IBM Miami and IBM Berlin~\cite{ibm_nighthawk_calibration}.
Table~\ref{tab:device_params} lists the coherence times, operational latencies, and syndrome extraction round duration (QEC round). To account for the unpublished reset latencies, we assume values from comparable processors. For Google Willow, we use the reset latency of Google Sycamore, whereas for IBM Miami and IBM Berlin, we use the duration of IBMQ Heron processors.

\begin{table}[htp]
%\vspace{-0.2in}
 \begin{center}
  \caption{Device parameters.}
 \setlength{\tabcolsep}{1.5mm}
 \renewcommand{\arraystretch}{1.0}
 \setcellgapes{0pt}\makegapedcells
 \label{tab:device_params}
 \begin{tabular}{|l||c|c|c|c|}
 \hline
 \multicolumn{1}{|c||}{Parameter}
 & \makecell{Google\\Sycamore} & \makecell{Google\\Willow}
 & \makecell{IBM\\Miami} & \makecell{IBM\\Berlin} \\
 \hline\hline
 $T_1$ (\textmu s)            &  20 &   68 &  301 &  331 \\ \hline
 $T_2$ (\textmu s)            &  30 &   89 &  232 &  270 \\ \hline
 1Q gate (ns)                 &  25 &   25 &   32 &   32 \\ \hline
 2Q gate (ns)                 &  34 &   42 &  161 &   76 \\ \hline
 Reset (ns)                   & 160 &  160 &   32 &   32 \\ \hline
 $L_\text{base}$ (ns) & 500 &  480 & 2400 & 2400 \\ \hline
 \hline
 QEC round (ns)               & 921 & 1108 & 3140 & 2800 \\ \hline
 \end{tabular}
 \end{center}
\end{table}
  
\subsection{Baseline and Error Model}
\label{methodology:baseline}

\noindent \textbf{Baseline:}
We use the rotated surface code with $L_\text{base}$ as each vendor's published calibration latency that minimizes measurement error rate ($p_\text{meas}$). We apply this latency uniformly to every syndrome extraction round. We evaluate code distances $d \in \{3, 5, 7, 9, 11, 13\}$ using Monte Carlo simulations.

\vspace{0.05in}
\noindent \textbf{Noise Model:}
We use a circuit-level depolarizing noise model with a uniform physical error rate $p$~\cite{raussendorf2007fault, wang2011surface, fowler2012surface, landahl2011fault, gidney2021fault}, typically used in QEC research~\cite{ravi2023better, chamberland2019fault, vittal2023astrea, vittal2024flag, gidney2025factor, knapen2026pinball, katsuda2024simulation}. We inject errors at seven stages of each syndrome extraction round:

\begin{enumerate}[leftmargin=0cm,itemindent=.5cm,labelwidth=\itemindent,labelsep=0cm,align=left, itemsep=0pt, listparindent=0cm]

\item \textbf{Single-qubit gates} (H, S, $\sqrt{X}$): A single-qubit depolarizing channel with error rate $0.1 p$.

\item \textbf{Two-qubit gates} (CZ): A two-qubit depolarizing channel with error rate $p$.

\item \textbf{Measurement}: Depolarizing channel with error rate $p_\text{meas}(L)$, which at the baseline $L_\text{base}$ equals $p$.

\item \textbf{Reset}: Depolarizing channel with error rate $p$.

\item \textbf{Decoherence}: At the beginning of each round, decoherence is injected as Pauli $X$, $Y$, and $Z$ errors with probabilities $p_X$, $p_Y$, and $p_Z$ as defined in Equation~\eqref{eq:pxy} and Equation~\eqref{eq:pz}. Under Pauli-twirling approximation~\cite{tomita2014low}, these probabilities depend on the qubit's idle time $t$ and coherence times $T_1$ and $T_2$:
 \begin{gather}
        p_X = p_Y = \tfrac{1}{4}\left(1 - e^{-t/T_1}\right) \label{eq:pxy}\\
        p_Z = \tfrac{1}{2}\left(1 - e^{-t/T_2}\right) - \tfrac{1}{4}\left(1 - e^{-t/T_1}\right) \label{eq:pz}
 \end{gather}

\item \textbf{Dynamical decoupling}: Dynamical decoupling (DD)~\cite{khodjasteh2005fault, souza2011robust, bylander2011noise, khodjasteh2007performance} suppresses idle errors by keeping idle qubits busy using a low-cost identity gate sequence (such as X--Y--X--Y or X--X or Y--Y) repeatedly while ensuring the quantum state remains unchanged. Google systems report coherence times already measured under DD ($T_{2,\mathrm{CPMG}}$~\cite{google2023suppressing, google2025quantum_supp}); so, we do not specifically apply DD gates.  
In contrast, IBM Miami and IBM Berlin report a Hahn-echo $T_2$ measured with a single refocusing pulse~\cite{ibm_dynamical_decoupling_hahn}. So, we apply one DD sequence every 500\,ns, corresponding to roughly five sequences per calibrated measurement. We apply a 
single-qubit depolarizing channel at an error rate of $0.1p$
for the single-qubit gates used for DD. 

\item \textbf{Idling}: Idling errors occur on inactive qubits during each two-qubit gate, at a rate of $0.1p$.

\end{enumerate}

%The measurement stage draws $p_\text{meas}(L)$ from the analytical model of Section~\ref{design:error_model}.
%We also evaluate \Dynamite{} where device error rates are non-uniform across qubits and drift between calibration cycles.

\begin{figure*}[!htp]
  \centering
  \includegraphics[width=\textwidth]{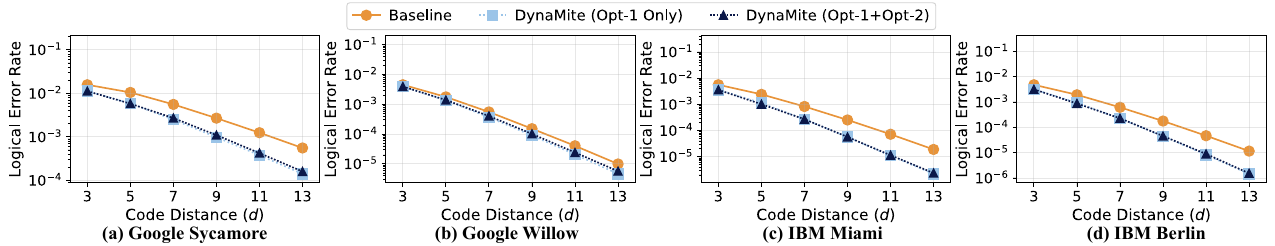}
  \vspace{-0.20in}
  \caption{LER versus code distance in a memory experiment ($p{=}0.1\%$, 100 million logical cycles).}
  \label{fig:ler_vs_d}
\end{figure*}

\begin{figure*}[!htp]
  \centering
  \includegraphics[width=\textwidth]{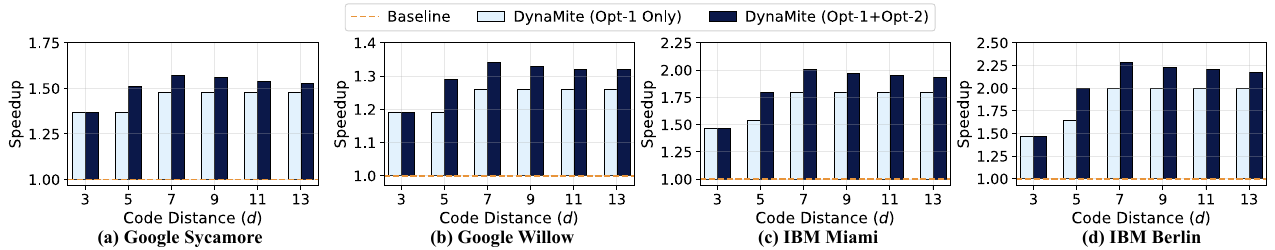}
  \vspace{-0.20in}
  \caption{Speedup versus code distance in a memory experiment ($p{=}0.1\%$, 100 million logical cycles).}
  \label{fig:speedup_vs_d}\vspace{-0.2in}
\end{figure*}

\subsection{Simulation Infrastructure}
\label{methodology:infrastructure}
\noindent \textbf{Frameworks:} We use Google's 
Stim framework~\cite{gidney2021stim} for circuit generation and error sampling, and PyMatching~\cite{Higgott2025sparseblossom}, a Minimum Weight Perfect Matching (MWPM) decoder that weights each error event by its probability.

\vspace{0.05in}
\noindent\review{\textbf{Adaptive sampling:} 
For \Dynamite{} (Opt-1+Opt-2), each trial selects a measurement latency per round based on syndromes from earlier rounds. 
After each $r^{th}$ round, the simulator computes $HW_{cum}$ of the trial, selects the latency $L$ of round $r{+}1$ with Equation~\ref{eq:find_latency} (matching \Dynamite{}'s selection), and injects errors in round $r{+}1^{th}$ for measurements, decoherence, and errors in DD sequences based on the selected latency. %Thus, the logical error reflects the measurement latencies used.
}

\vspace{0.05in}
\noindent \textbf{Experiment types:}
We evaluate \Dynamite{} across two Monte Carlo regimes.
(i) \textit{Memory experiments} at $p{=}0.1\%$ ($100$ million trials, $d \in \{3, 5, 7, 9, 11, 13\}$) test logical-qubit preservation, and 
(ii) \textit{Better-hardware experiments} raise the coherence times at $p{=}0.1\%$ ($100$ million trials, $d \in \{7, 9, 11\}$) to evaluate \Dynamite{}'s effectiveness on future hardware.
We cannot evaluate larger distances at such low error rates because the required number of Monte Carlo trials for statistical significance becomes infeasible. Our approach is consistent with prior works that use similar distances and setup~\cite{alavisamani2024promatch, vittal2023astrea, toshio2025decoder}.

\subsection{Figure-of-Merit}
\label{methodology:metrics}
  
We evaluate \Dynamite{} using two key metrics.

\vspace{0.05in}
\noindent \textbf{Relative LER: }
The LER of a given configuration relative to the baseline that uses $L_\text{base}$ for all qubit measurements.
A value of 1 means no impact on LER; values above it reflect a higher LER, whereas values below it a lower LER.%\textit{Ideally, a lower Relative LER with values up to 1 is desirable because it indicates no impact on error tolerance. }

\vspace{0.05in}
\noindent \textbf{Speedup: }
We compute the speedup as the ratio of the baseline LCT to that of \Dynamite{}.
For example, a speedup of $1.43\times$ means the logical cycle completes in 70\% of the baseline time.
\textit{Ideally, lower Relative LER and higher speedup are desirable.}

\section{Results}
\label{sec:evaluation}

\subsection{Impact on Logical Performance and Physical Overhead}
\label{sec:cross_device}

Figures~\ref{fig:ler_vs_d} and \ref{fig:speedup_vs_d} show the performance of \Dynamite{} across four systems.
%Figure~\ref{fig:ler_speedup_vs_d} shows the performance of \Dynamite{} across four systems.
\Dynamite{} with Opt-1 Only achieves a speedup of  $1.56\times$ on average and up to $2.00\times$. It offers a Relative LER of $0.44\times$ on average and up to $0.12\times$ in the best case (for IBM Miami $d{=}13$). \Dynamite{} with both Opt-1+Opt-2 that selects the latency dynamically for each QEC round based on real-time error budget tracking improves speedup to $1.69\times$ on average and up to $2.29\times$; and offers a Relative LER of $0.46\times$ on average and up to $0.13\times$ compared to the baseline.

The efficacy of Opt-2 depends on range of latencies available below $L_\text{opt1}$. For example, on Google Willow for distance-7, $L_\text{opt1}$ is already low (250\,ns), leaving very little room (50\,ns to 250\,ns only) for Opt-2. As a result, \Dynamite{} (Opt-1+Opt-2) improves speedup from $1.26\times$ (with Opt-1 alone) to only $1.34\times$. In contrast, on IBM Berlin for distance-7, the $L_\text{opt1}$ is 1000\,ns, which allows Opt-2 to select latencies from a wider range (100\,ns to 1000\,ns). Consequently, \Dynamite{} (Opt-1+Opt-2) improves speedup from $2.00\times$ (with Opt-1 alone) to $2.29\times$. \Dynamite{} (Opt-1+Opt-2) consistently outperforms \Dynamite{} (with Opt-1 alone) across all code distances.

As \Dynamite{} achieves Relative LER always below 1.0, it requires lower redundancy (or physical overheads) to achieve the same target LER. For example, on IBM Berlin, a target LER of $4.65\times10^{-5}$ requires $d{=}11$ (241 qubits) for the baseline but only $d{=}9$ (161 qubits) for \Dynamite{}, \textbf{yielding a 33.2\% qubit reduction}. Moreover, the Relative LER gains increase with code distance. Hence, the physical overhead reduction benefits are even more significant in practical applications.

\subsection{\review{Opt-1+Opt-2 at Large Distances and Program Runtime}}
\label{eval:large_distance_program_runtime}

\begin{figure}[htp]
   \centering
   \includegraphics[width=\columnwidth]{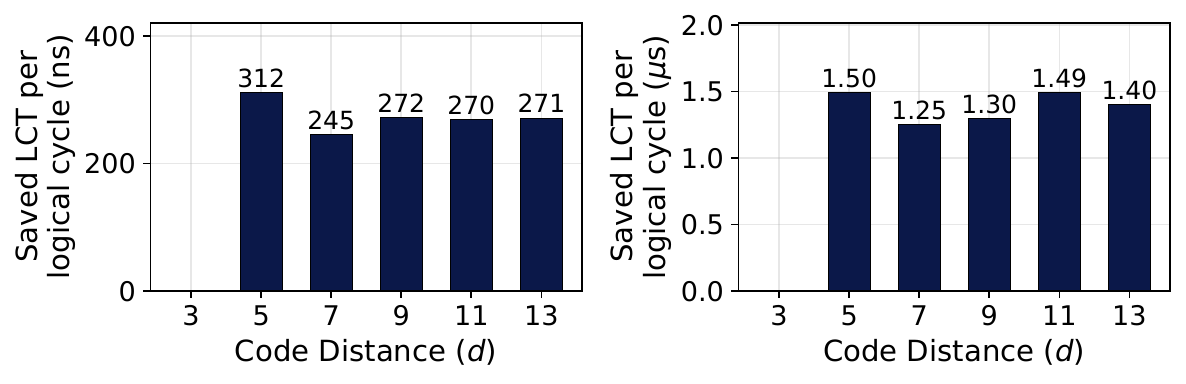}
   \vspace{-0.25in}
   \caption{\review{Saved LCT per logical cycle of \Dynamite{} (Opt-1+Opt-2) over \Dynamite{} (Opt-1 Only) versus code distance.}}% in a memory experiment ($p{=}0.1\%$, 100 million logical cycles).}}
   \label{fig:saved_time_distance}
\end{figure}

\review{Figure~\ref{fig:saved_time_distance} shows LCT savings by \Dynamite{} (Opt-1+Opt-2) per logical cycle over \Dynamite{} (Opt-1 Only). %For $d = 7$ to 13 the time saved stays the same or grows (245 to 272\,ns on Google Sycamore and 1.25 to 1.49~$\mu$s on IBM Berlin).
%The relative gain over Opt-1 Only falls with $d$ only because the LCT, $d$ times the QEC round, grows with $d$.
The savings do not shrink with $d$, so there is no distance when gain vanishes.
}
%\hl{At $d = 13$ the LER stays at the \Dynamite{} (Opt-1 Only) level: the ratio is 1.06 to $1.24\times$, whereas one code-distance step lowers the LER by 2.7 to $5.6\times$.}

\review{To show that even nominal savings in LCT can translate to a big difference, we compute the wall-clock time (WCT) = $N \times$ LCT, where $N$ is the number of logical cycles, from our $d = 13$ results for three FTQC workloads, which spend one logical cycle per T gate: 2048-bit factoring~\cite{gidney2021_rsa_factorization} ($N = 6.1\times10^9$ T gates~\cite{litinski2022active}) and cytochrome P450 with THC ($N = 3.1\times10^{10}$) and BLISS-THC ($N = 6.9\times10^9$)~\cite{caesura2025faster}. Table~\ref{tab:wall_clock_time} lists the results.
\Dynamite{} (Opt-1+Opt-2) yields additional 0.5, 2.3, and 0.5 hours savings on Google Sycamore and 2.4, 12.1, and 2.7 hours on IBM Berlin. The limited availability of quantum computers makes it necessary to improve system throughput~\cite{das2019case, liu2021qucloud, niu2023enabling, wakizaka2025online, ravi2021quantum, meng2024qoncord, ravi2021adaptive} by reducing per program runtimes.
%This saving is $N$ times the per-cycle saving, which stays flat for $d = 7$ to 13, so it is expected to carry over to larger distances.
}

\begin{table}[!htp]
  \begin{center}
  \caption{\review{Wall-clock time (hours) at $d = 13$.}}
  \setlength{\tabcolsep}{2pt}
  \renewcommand{\arraystretch}{1.0}
  \label{tab:wall_clock_time}
  \small
  \begin{tabular}{|c|c||c|c|c|}
    \hline
    \makecell[c]{Workload} &
    \makecell[c]{Device} &
    \makecell[c]{Baseline} &
    \makecell[c]{Opt-1\\Only} &
    \makecell[c]{Opt-1+\\Opt-2} \\
    \hline\hline
    \multirow{2}{*}{\shortstack{2048-bit\\factoring}}
      & Google Sycamore & 20.3 & 13.7 & 13.2 \\
    \cline{2-5}
      & IBM Berlin & 61.7 & 30.8 & 28.5 \\
    \hline
    \multirow{2}{*}{P450 (THC)}
      & Google Sycamore & 103.6 & 69.8 & 67.5 \\
    \cline{2-5}
      & IBM Berlin & 314.9 & 157.5 & 145.3 \\
    \hline
    \multirow{2}{*}{\shortstack{P450\\(BLISS-THC)}}
      & Google Sycamore & 22.8 & 15.4 & 14.9 \\
    \cline{2-5}
      & IBM Berlin & 69.3 & 34.7 & 32.0 \\
    \hline
  \end{tabular}
  \end{center}
\end{table}

\review{
Larger codes provide more opportunities to \Dynamite{} because they have more rounds per logical cycle and we expect physical error rates ($p$) to go down to run such   
  codes in practice. Syndrome vector HW scales with $p$. For example, a code of distance $d$ has $\sim d \times(3d^2)$ potential fault locations per logical cycle and each may fail with probability $p$. So, the average syndrome vector HW is $\sim6d^3p$~\cite{das2022afs} because each failure flips at most 2 bits. At $d$=17 and $p{=}10^{-3}$, average syndrome HW is $\sim30$, which   
  reduces to $\sim3$ when $p$ falls to $10^{-4}$. Thus, at lower $p$, more logical cycles result in low HW. While we cannot simulate large $d$ at low $p$, we  
  observe that for $d$=7 at $p{=}10^{-4}$ ($T_1/T_2 = 150/200\mu s$), 87\% of logical cycles have HW below error budget compared to 61\% at $p=10^{-3}$ on Google Sycamore. }

\subsection{Understanding HW Distribution and Error Chain Changes}

\begin{figure}[htp]                               
   \centering
   \includegraphics[width=\columnwidth]{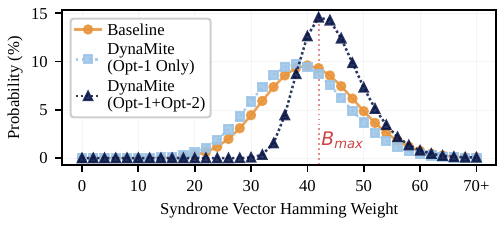}
   \vspace{-0.25in}
   \caption{Syndrome vector HW distribution for $d{=}11$ on Google Sycamore (100 million logical cycles, $p{=}0.1\%$). \Dynamite{} (Opt-1+Opt-2) shifts the distribution toward $B_{\text{max}}$ by exploiting the underutilized error budget.}
   \label{fig:eval_hw_dist}
\end{figure}

Figure~\ref{fig:eval_hw_dist} shows the syndrome vector HW distribution of a distance-11 surface code on Google Sycamore. Opt-1 does not alter the HW distribution significantly, yet the LER is lower because it has different conditional logical failure rates than the baseline. Opt-2 allows more effective utilization of the error budget. The number of logical cycles with very low HW is reduced, while more have HW closer to $B_\text{max}=42$. For example, in the \Dynamite{} (Opt-1), 17\% and 53\% of logical cycles end with HWs between 0--30 and 31--42, respectively. 
In contrast, in \Dynamite{} (Opt-1+Opt-2), 0\% and 42\% of logical cycles end with HWs between 0--30 and 31--42, respectively. 

%how \Dynamite{} (Opt-1+Opt-2) spends the underutilized error budget: for $d{=}11$ on Google Sycamore, running each round at a faster latency raises the syndrome vector HW and shifts the distribution toward $B_\text{max}$, so fewer cycles finish with HW$<B_\text{max}$.

\begin{figure}[htp]
 \centering
 \includegraphics[width=\columnwidth]{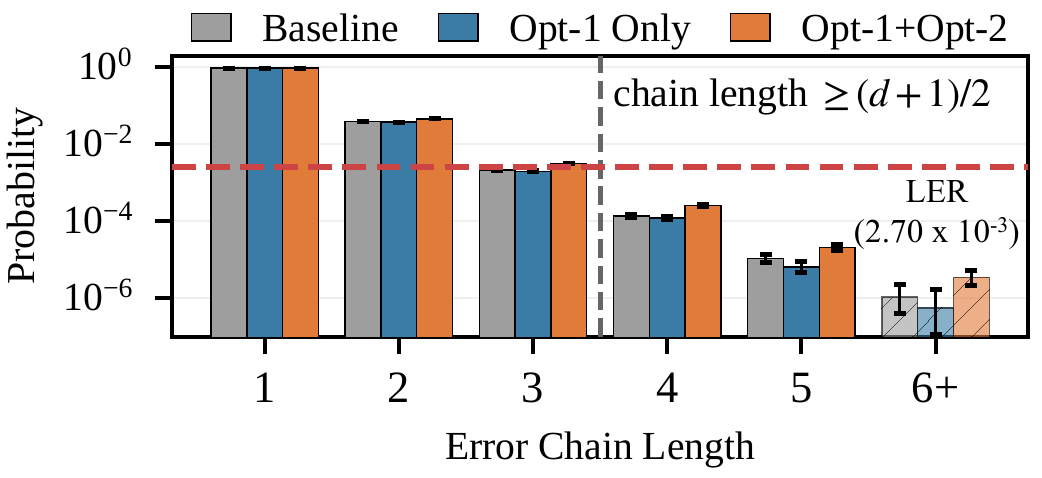}
 \vspace{-0.25in}
     \caption{Error chain length distribution for $d{=}7$ on Google Sycamore (1 million logical cycles, $p{=}0.1\%$). \Dynamite{} (Opt-1+Opt-2) encounters more error chains, yet almost all of them are short. The dashed line marks its LER.}
 \label{fig:error_chain_dynamite}
\end{figure}

Figure~\ref{fig:error_chain_dynamite} shows the error chain length distribution of a distance-7 surface code on Google Sycamore. \Dynamite{} (Opt-1 Only) does not alter the distribution significantly.
\Dynamite{} (Opt-1+Opt-2) encounters more error chains because faster measurements add errors, yet almost all of them are short.
For example, \Dynamite{} (Opt-1 Only) and \Dynamite{} (Opt-1+Opt-2) encounter 5.32 and 6.61 error chains per logical cycle, respectively, and 99.97\% of the chains in \Dynamite{} (Opt-1+Opt-2) contain at most three errors. Thus, the LER increases by only $1.09\times$, from $2.49{\times}10^{-3}$ to $2.70{\times}10^{-3}$.

\subsection{Logical Performance for Better Hardware}
\label{eval:better_hw}

\begin{figure}[htp]                               
   \centering
   %\vspace{-0.20in}
   \includegraphics[width=\columnwidth]{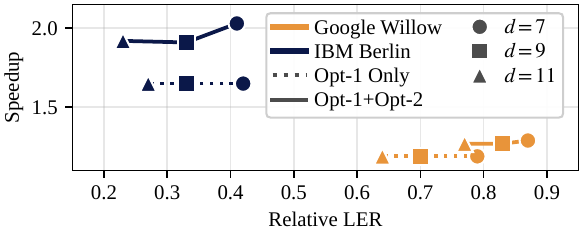}
   \vspace{-0.25in}
   \caption{Speedup versus Relative LER with reduced hardware error rates. \Dynamite{} remains effective even when hardware characteristics improve.}
   \label{fig:better_hw}
\end{figure}

We evaluate \Dynamite{} for improved coherence times to test its efficacy as devices improve ($T_1$ and $T_2$ go up to $150$\,\textmu s and $200$\,\textmu s for Google Willow; and $500$\,\textmu s and $400$\,\textmu s for IBM Berlin) and $p{=}0.1\%$. Figure~\ref{fig:better_hw} shows the Relative LER and speedup. \Dynamite{} (Opt-1 Only) achieves an average speedup of $1.42\times$ and a Relative LER of $0.53\times$. \Dynamite{} (Opt-1+Opt-2) raises the average speedup to $1.62\times$ at a Relative LER of $0.57\times$, confirming that \Dynamite{} remains effective even as the error-rates improve in the future.

% IBM Berlin Future
% d=9 -> EHW: 9.57, Bmax=13 -> fraction: 26.4%
% d=11 -> EHW:17.71, Bmax=21 -> fraction: 15.7%

% memory experiment: logical observable <- 'xor' of data qubits in one of the edge on surface code
% stability experiment: logical observablce <- 'xor' of stabilizers across multi time (multi rounds)

\subsection{Impact of Non-Uniform Error Rates and Drift}
\label{sec:sensitivity}

Real devices exhibit non-uniform error rates~\cite{ibm_nighthawk_calibration, google2025quantum_supp, tannu2019not} and drift over time~\cite{klimov2018fluctuations}. To evaluate efficacy of \Dynamite{} in these circumstances, we adjust the noise model to introduce non-uniformity by drawing each qubit's $T_1/T_2$, its single-qubit, two-qubit, measurement, and reset error rates from Gaussian distributions. For $T_1/T_2$, we use a distribution centered around the device mean $T_1/T_2$ and select values from $[0.5\times, 1.5\times]$ of the mean values. For single-qubit errors, we consider a distribution with a mean of $0.1p$ and values from $[0.05p, 0.15p]$. For two-qubit, measurement, and reset operations, we consider distributions centered around a mean error rate of $p$ and select values from $[0.5p, 1.5p]$. Other errors inherit this variation: decoherence follows each $T_1/T_2$, dynamical-decoupling and spectator-idling errors use each qubit's single-qubit rate ($0.1p$ nominal, in $[0.05p, 0.15p]$). For drift, we perturb per-qubit rates daily over three days.
Table~\ref{tab:calibration} shows that \Dynamite{} (Opt-1+Opt-2) reduces Relative LER to $0.38\times$ on average and achieves $1.77\times$ speedup on average.
Speedup varies only marginally across days (within $0.23\%$). %Although \Dynamite{} (Opt-1+Opt-2) selects latencies online, it operates based on aggregate error budget and is insensitive small drifts, so we report a single representative value. 
\Dynamite{} stays effective because it is driven by aggregate error budget rather than per qubit error rate, leaving it robust to per-qubit variation and drifts. Also, systems are largely stable over short periods and exhibit variability over extended time (weeks 
or months)~\cite{dasgupta2021stability} and newer systems are more stable to drift. \review{For example, standard deviation of two-qubit error rates over 30 days is only 0.86\% on recent IBM Heron~\cite{ibm_heron_calibration} compared to 3.02\% on IBMQ Tokyo from 2018~\cite{tannu2019not}.}

% Google Sycamore: (1.5561-1.5556)/1.5561 = 0.032%
% Google Willow:   (1.3293-1.3290)/1.3293 = 0.023%
% IBM Miami:       (1.9705-1.9690)/1.9705 = 0.076%
% IBM Berlin:      (2.2282-2.2232)/2.2282 = 0.224%

\begin{table}[!htp]
      \begin{center}
      \caption{\Dynamite{} (Opt-1+Opt-2) under non-uniform per qubit error rates and drift ($d=9$, $p=0.1\%$, 100 million logical cycles).}
      \setlength{\tabcolsep}{1.5mm}
      \renewcommand{\arraystretch}{1.0}
      \label{tab:calibration}
      \begin{tabular}{|l||c|c|c||c|}
        \hline
        \multirow{2}{*}{Device} & \multicolumn{3}{c||}{Relative LER} & \multirow{2}{*}{Speedup} \\
        \cline{2-4}
         & Day 1 & Day 2 & Day 3 & \\
        \hline\hline
        Google Sycamore & 0.41 & 0.41 & 0.42 & 1.56 \\
        \hline
        Google Willow   & 0.65 & 0.65 & 0.64 & 1.33 \\
        \hline
        IBM Miami       & 0.22 & 0.22 & 0.22 & 1.97 \\
        \hline
        IBM Berlin      & 0.25 & 0.25 & 0.25 & 2.23 \\
        \hline
      \end{tabular}
      \end{center}
\end{table}

\subsection{Sensitivity to Noise Parameters and Drift}
\label{sec:ac_sensitivity}

\Dynamite{} derives $L_{\text{opt1}}$ from two parameters, the signal-to-noise ratio growth rate $a$ and relaxation prefactor $c$. 
Any miscalibration may yield sub-optimal  $L_{\text{opt1}}$ and degrade LER. 
To evaluate this, we set each parameter to $\{0.7\times, 1.0\times, 1.3\times\}$ its calibrated value, forming a $3{\times}3$ grid of nine $(a,c)$ combinations spanning up to $30\%$ miscalibration, and evaluate \Dynamite{} (Opt-1 Only). We compare two policies: (1)~\emph{Fixed} that keeps $L_{\text{opt1}}$ unchanged, and (2)~\emph{Adaptive} that re-derives $L_{\text{opt1}}$ for each $(a,c)$.
Table~\ref{tab:ac_sensitivity} reports Relative LER that stays within $0.12$--$0.86\times$. \Dynamite{} uses adaptive policy by default which outperforms fixed policy. Keeping $L_{\text{opt1}}$ fixed is far less robust. For example, on IBM Berlin, Relative LER degrades to $1.69\times$ when $a$ is reduced by $30\%$, because $L_{\text{opt1}}$ shifts to $1500$\,ns rather than $1000$\,ns at optimal parameter values, leaving fixed policy suboptimal.
\review{$L_{opt1}$ is mainly sensitive to $a$: a 30\% lower $a$ moves it up by 100 to 150\,ns on Google devices and by 500\,ns on IBM machines, a 30\% higher $a$ moves it by at most 100\,ns, $c$ has no visible effect. Re-deriving $L_{opt1}$ resolves this issue, which is a cheap offline table lookup}. So, \Dynamite{} obtains this robustness at no runtime cost.

\begin{table}[!htp]
      \begin{center}
      \caption{\Dynamite{} (Opt-1 Only) under $\pm30\%$ miscalibration of $(a,c)$ \\($d=9$, $p=0.1\%$, 10 million logical cycles).}
      \setlength{\tabcolsep}{1mm}
      \renewcommand{\arraystretch}{1.0}
      \label{tab:ac_sensitivity}
      \small
      \begin{tabular}{|l|c|c|}
      \hline
      \multirow{2}{*}{Device} & \multicolumn{2}{c|}{Relative LER range} \\
      \cline{2-3}
       & Fixed $L_{\text{opt1}}$ & \Dynamite{}'s rule \\
      \hline\hline
      Google Sycamore & 0.26--0.86 & 0.26--0.59 \\
      \hline
      Google Willow   & 0.47--1.20 & 0.47--0.86 \\
      \hline
      IBM Miami       & 0.12--1.29 & 0.12--0.47 \\
      \hline
      IBM Berlin      & 0.14--1.69 & 0.14--0.50 \\
      \hline
      \end{tabular}
      \end{center}
  \end{table}

\subsection{\review{Impact of a Complex Readout Model}}
\label{sec:complex_readout_model}

\begin{figure}[htp]
   \centering
   \includegraphics[width=\columnwidth]{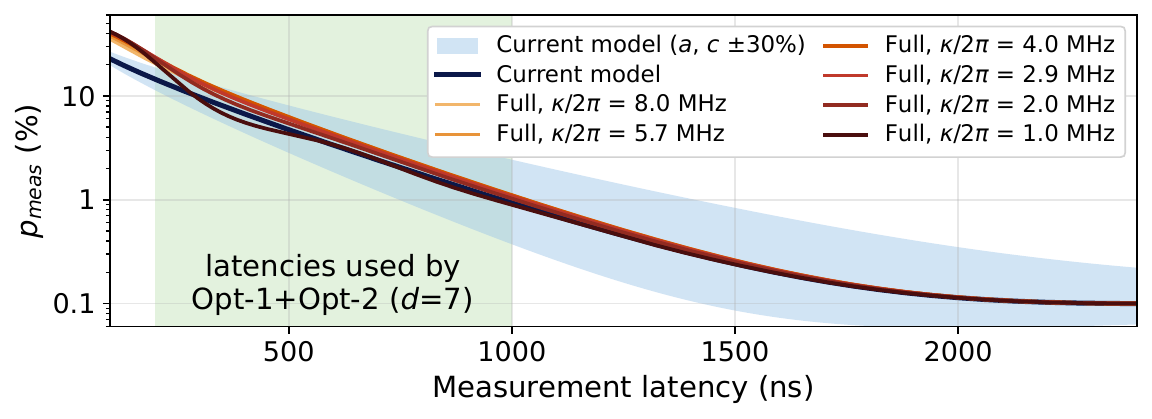}
   \vspace{-0.25in}
   \caption{{Readout error ($p_{meas}$) versus latency on IBM Berlin for current and full model (separation, relaxation, ring-up, linewidth, and photon number).}}% at linewidths $\kappa/2\pi$ from 1 to 8\,MHz, all calibrated to the same conditions. The band shows a $\pm$30\% change in $a$ and $c$; the green region marks the latencies \Dynamite{} (Opt-1+Opt-2) uses at $d$=7. The resonator parameters are assumed (a dispersive shift $2\chi/2\pi$ of 4\,MHz), as those of IBM Berlin are not public.}}
   \label{fig:full_readout}
\end{figure}

\review{
%The readout model enters \Dynamite{} only through the measurement error at each latency, so a more detailed model changes the entries of the LLT, and with them $L_{opt1}$ and the safe thresholds $\tau_r$, but not the policy.
Our evaluations are based on the two-part error model comprising separation and relaxation errors (Equation~\ref{eq:meas_error_1} below).

\begin{equation}
    \label{eq:meas_error_1}
        p_{\text{meas}}(t) = \underbrace{\tfrac{1}{2}\,
        \text{erfc}\!\left(\tfrac{\sqrt{a \cdot t}}{2}\right)}_{E_{\text{sep}}(t)}
        + \underbrace{c \cdot \left(1 - e^{-t / T_1}\right)}_{E_{\text{relax}}(t)}
\end{equation}

% follows the dispersive readout model of prior device-level studies~\cite{krantz2019quantum, bengtsson2024model}, and the soft-measurement model of Pattison et al.~\cite{pattison2021_softInfo} has a similar structure: a separation error that falls with the measurement latency and a relaxation error that grows with it.
Resonator parameters like linewidth $\kappa = \omega_r/Q$ (resonator frequency $\omega_r$ over quality factor $Q$) and readout photon number $\bar{n}$ set how fast the signal-to-noise ratio grows with time, so they are already factored in parameter $a$ as shown in Equation~\ref{eq:meas_error_full}, where $2\chi$ is the dispersive shift and $\eta$ is measurement efficiency~\cite{bengtsson2024model}.
Ring-up enters the separation error through $T_{\text{eff}}$, whereas ring-down does not affect it.}
\review{
%The photon number, linewidth, and ring-up enter only the separation error (Equation~\ref{eq:meas_error_full}), where $2\chi$ is the dispersive shift and $\eta$ the measurement efficiency~\cite{bengtsson2024model}:
\begin{equation}
  \label{eq:meas_error_full}
  E_{\text{sep}}(t) = \tfrac{1}{2}\,\text{erfc}\!\left(\tfrac{\sqrt{a \cdot T_{\text{eff}}(t)}}{2}\right), \quad
  a = \frac{8\,\eta\,\kappa\,\chi^2\,\bar{n}}{(\kappa/2)^2 + \chi^2}
\end{equation}

%Calibration fits $a$ directly, so $\bar{n}$, $\eta$, and $\chi$ need not be known.
The linewidth $\kappa$ also sets the ring-up: the resonator starts empty, so its responses to the two qubit states separate over about $2/\kappa$~\cite{bengtsson2024model}.
The effective time $T_{\text{eff}}(t) = \int_0^t s(t')^2\,dt'$ captures this, where $s(t)$ is the distance between resonator signals for the two states, normalized to 1 once the resonator is full; Equation~\ref{eq:meas_error_1} is the case $s=1$. We vary $\kappa/2\pi$ from 1 to 8\,MHz and show that fidelity versus latency trade-offs remain comparable to our two-component model. Also note that most of these variations are already captured in our sensitivity studies in Section~\ref{sec:ac_sensitivity}, where we show that \Dynamite{} remains effective even if parameters $a$ and $c$ vary within $\pm$30\% (which is confirmed by the blue region in Figure~\ref{fig:full_readout}). 
Ring-down~\cite{mcclure2016rapid} does not affect $p_{\text{meas}}$ but adds the same time to every round of both the baseline and \Dynamite{}; with an active reset pulse that empties the resonator in about $2/\kappa$~\cite{mcclure2016rapid}, the speedup at $d$=7 on IBM Berlin stays at 2.03 to 2.25$\times$ for these linewidths.}

\noindent\review{\textbf{Model validation against real device data:} 
We compare the measurement error rates 
predicted by our two-component model against raw traces available from the discriminator paper by Maurya et al. from ISCA 2023~\cite{maurya2023scaling}. Table~\ref{tab:readout_validation} shows that the readout traces follow similar latency versus fidelity trends and our predictions closely match the actual device. }

\begin{table}[htp]
    \centering\small\setlength{\tabcolsep}{2pt}
    \renewcommand{\arraystretch}{1.0}
    \caption{{Measured and predicted $p_{meas}$ of a five-qubit device~\cite{maurya2023scaling}.}}
    \label{tab:readout_validation}
    \begin{tabular}{|c|c|c|c|c|c|c|}
    \hline
    \multirow{2}{*}{Qubit} & \multicolumn{3}{c|}{Measured $p_{meas}$} & \multicolumn{3}{c|}{Predicted $p_{meas}$} \\
    \cline{2-7}
    & 1\,$\mu$s & 750\,ns & 500\,ns & 1\,$\mu$s & 750\,ns & 500\,ns \\
    \hline\hline
    1 & 1.5\% & 4.9\% & 37.1\% (\textit{unusually high}) & 3.1\% & 5.3\% & 9.3\% \\
    2 & 24.6\% & 25.8\% & 29.2\% & 24.5\% & 26.2\% & 29.0\% \\
    3 & 3.4\% & 4.5\% & 9.0\% & 3.3\% & 4.9\% & 8.4\% \\
    4 & 3.8\% & 4.2\% & 7.1\% & 3.7\% & 4.4\% & 6.8\% \\
    5 & 1.1\% & 1.3\% & 2.3\% & 1.1\% & 1.3\% & 2.3\% \\
    \hline
    \end{tabular}
\end{table}

\subsection{Implementation Costs}
\label{eval:hardware}

\noindent\textbf{Latency:}
\review{\Dynamite{} \textit{only} changes the length of the next readout pulse. Discrimination and syndrome transfers happen in parallel for all readouts and take same time in the baseline and \Dynamite{}. 
%First, the readout pulse, discrimination, and syndrome transfer to the decoder run every round, in parallel for every ancilla.
%Second, each readout pulse and its discrimination threshold are calibrated offline, with no retraining at run time.
The additional serial steps are HW aggregation (HW agg.) and comparison against $\tau_r$ to decide the next round readout latency.}~
To select the latency for round $k$, \Dynamite{} must observe the previous round's measurement, update the syndrome vector HW, and decide before the current round's measurement is scheduled.
The budget for that decision is the sum of a reset (from previous round) and four CNOTs (from current round): \review{296\,ns on Sycamore, 328\,ns on Willow, 336\,ns on Berlin, and 676\,ns on Miami.}~The hardware involves a subtraction (budget update), a comparison against $\tau_k$, and table lookup.
\review{We synthesize \Dynamite{} for $d{=}13$ using Xilinx Vivado to target a commercial Zynq UltraScale+ FPGA at 200\,MHz.
With a register between the two steps (2.2 and 4.3\,ns) as shown in Table~\ref{tab:hardware}, each fits in one clock cycle, so \Dynamite{} selects the measurement latency in two clock cycles (10\,ns).
This leaves 286 to 666\,ns for discrimination and syndrome transfer. On the dynamic control hardware of Ryan et al.~\cite{ryan2017hardware}, discrimination takes 88\,ns and syndrome transfer takes 10\,ns, so their 98\,ns total fits within the budget.}

\noindent\textbf{Hardware overheads:} 
\review{\Dynamite{} uses 428 LUTs (0.16\%), 592 registers (0.11\%), and no BRAM or DSP for $d{=}13$, staying under 0.2\% of the FPGA, as shown in Table~\ref{tab:hardware}}. %This lightweight footprint allows seamless adoption in the control processors of existing quantum processors.

\begin{table}[htp]
    \centering\small\setlength{\tabcolsep}{2pt}
    \renewcommand{\arraystretch}{1.0}
    \caption{FPGA synthesis results for \Dynamite{} at $d{=}13$.}
    \label{tab:hardware}
    \begin{tabular}{|c|c|c|c|c|c|}
    \hline
    LUT & FF & BRAM & DSP & HW agg. & Decision \\
    \hline\hline
    428 (0.16\%) & 592 (0.11\%) & 0 & 0 & 2.2\,ns & 4.3\,ns \\
    \hline
    \end{tabular}
\end{table}

\noindent\review{\textbf{Decoder costs:} \Dynamite{} \textit{only} changes decoder weights, but not matching graph, which is independent of measurement latency. So, the decoding graph is never re-built. Instead, for each round, a mux selects precomputed 8-bit weights~\cite{vittal2023astrea} for the latency-dependent error locations in the graph. Extra memory required to store additional weights is (distinct latencies-1)$\times d\times$ (error locations per round)$\times$8\,bit, with about $4d^2$ error locations per round on Google devices (ancilla measurement and data-qubit idling) and $7d^2$ on IBM devices (including dynamical decoupling data), resulting in at most 107.7\,KB at $d{=}13$ (Table~\ref{tab:decoder_mem}).
Weight selection adds no decoding time: all weights stay in memory, and the mux select is issued when a round's latency is chosen, before syndrome arrives.}
%Fourth, decoding keeps up in real time: at $d{=}13$, the average LCT of \Dynamite{} (Opt-1+Opt-2) is 7.8\,$\mu$s on Google Sycamore and 16.8\,$\mu$s on IBM Berlin, at least nine times the 0.8\,$\mu$s average decoding latency~\cite{wu2025micro}.}

\begin{table}[!htp]
    \begin{center}
    \caption{{Additional decoder weight memory of
    \Dynamite{} (Opt-1+Opt-2) over \Dynamite{} (Opt-1 Only),
    ranges over the four devices.}}
    \setlength{\tabcolsep}{1.4mm}
    \renewcommand{\arraystretch}{1.1}
    \label{tab:decoder_mem}
    \small
    \begin{tabular}{|l|c|c|c|c|c|}
      \hline
      $d$ & 5 & 7 & 9 & 11 & 13 \\
      \hline\hline
      \makecell[l]{Error locations per\\round (Google / IBM)} &
      \makecell[c]{100/\\175} & \makecell[c]{196/\\343} & \makecell[c]{324/\\567} &
      \makecell[c]{484/\\847} & \makecell[c]{676/\\1,183} \\
      \hline
      Distinct latencies & 3 to 4 & 4 to 6 & 4 to 8 & 4 to 8 & 4 to 8 \\
      \hline
      \makecell[l]{Additional\\memory (KB)} &
      \makecell[c]{1.0 to\\1.8} & \makecell[c]{4.1 to\\12.0} & \makecell[c]{8.7 to\\35.7} &
      \makecell[c]{16.0 to\\65.2} & \makecell[c]{26.4 to\\107.7} \\
      \hline
    \end{tabular}
    \end{center}
\end{table}

\noindent
\phantomsection
\textbf{Calibration Overheads:}
\label{sec:cali_overhead}
These arise from two sources. First, \Dynamite{} must profile the error rate per latency, which it can measure directly using randomized benchmarking (${\sim}10^4$ circuit executions per latency, tens of milliseconds), or more cheaply by decomposing a single larger-latency signal trace into shorter segments (for example, a 200\,ns trace from Willow's 480\,ns calibrated $L_\text{base}$), which eliminates the extra per-latency execution. Alternatively, \Dynamite{} can use the analytical model and $T_1$/$T_2$/$p_\text{meas}$ values from calibrations to populate the LLT. 
Second, \Dynamite{} must profile the mean Hamming weight of syndrome vectors. Estimating this requires about 1K syndrome vectors, which on superconducting systems (where a logical cycle takes $d$ microseconds) takes about $d$ milliseconds, negligible compared to routine calibrations.

\begin{figure}[htp]
     \centering
     \includegraphics[width=\columnwidth]{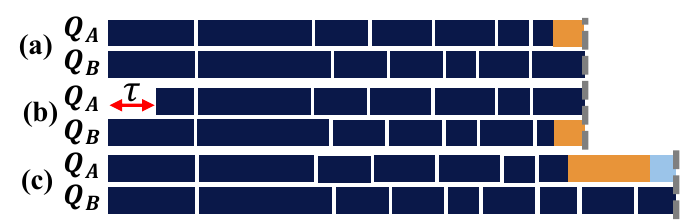}
     \vspace{-0.25in}
     \caption{{Synchronization of two logical qubits ($Q_A$, $Q_B$) under \Dynamite{} (Opt-1+Opt-2). Dark blocks denote QEC rounds with variable lengths, orange shows how \Dynamite{} lengthens the final measurement (up to $L_{base}$), light blue marks idle time, and the dashed line marks end of logical cycle. (a) same distance with aligned start (Exp.~1), (b) same distance with $Q_A$ starting later by $\tau$ (Exp.~2), and (c) $Q_A$ at $d$=7 and $Q_B$ at $d$=9 with aligned start (Exp.~3).}}
     \label{fig:sync_policy}
\end{figure}

\review{
\subsection{Impact on Synchronization between Logical Qubits}
\label{sec:logical_qubits_synchronization}
Lattice surgery requires the logical cycles of the participating qubit patches to be synchronized before joint parity checks can be performed. Although \Dynamite{} (Opt-1+Opt-2) may cause the logical cycles of two qubits to be desynchronized as they may use different LCTs, the impact on performance is negligible due to two reasons. \textit{First}, only active qubits require synchronization whereas most qubits remain idle. For example, in a 2048-bit factoring program, 97.7\% of the qubits remain idle, and only 2.3\% of logical qubits are active in a logical cycle on average. \textit{Second}, for two qubits requiring synchronization, \Dynamite{} simply adjusts the latency of the last QEC round of each qubit so that their logical cycle endpoints align; the qubit that finishes earlier lengthens its final round measurement (up to $L_{base}$) so that the cycles for both qubits end at the same time. 
To evaluate the impact on performance, we conduct additional experiments for three scenarios: (a) Exp-1: when two qubits $Q_A$ and $Q_B$ start their logical cycles at the same time but one finishes earlier and must extend its last round, (b) Exp-2: when one qubit ($Q_A$) starts its logical cycle later by $\tau$ than the other qubit ($Q_B$) to represent a scenario where the previous cycles were desynchronized, (c) Exp-3: when two qubits of different distances are used.
Figure~\ref{fig:sync_policy} illustrates the three scenarios.}

\begin{figure}[htp]
       \centering
       \includegraphics[width=\columnwidth]{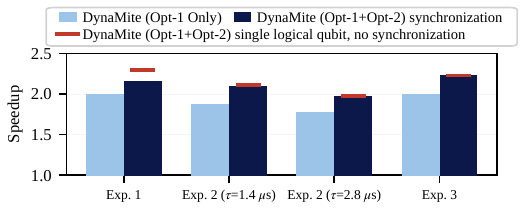}
       \vspace{-0.25in}
  \caption{{Speedup of two synchronized logical qubits.}}
      \label{fig:lct_sync}
\end{figure}

\noindent\review{\textbf{Results:} Figure~\ref{fig:lct_sync} shows that the synchronized pair stays faster than \Dynamite{} (Opt-1 Only) consistently. For Exp.~1, Exp.~2 at $\tau$ of 1.4 and 2.8\,$\mu$s, and Exp.~3, the pair reaches 2.17$\times$, 2.09$\times$, 1.97$\times$, and 2.23$\times$ over baseline, against 2.00$\times$, 1.88$\times$, 1.78$\times$, and 2.00$\times$ for \Dynamite{} (Opt-1 Only). This shows that even when \textbf{\textit{\Dynamite{} adjusts the last round to synchronize two qubits, the impact on LCT is negligible}}.} %, and stays close to a single logical qubit.}

\noindent\review{\textbf{Calibrating on hardware:} We find $L_{opt1}$ by simulating QEC at every candidate latency with the measurement error of Equation~\ref{eq:meas_error}; the same sweep can run on hardware, and the latency lookup table (LLT) can likewise be filled with measured values.
Reliably measuring a logical error rate near $10^{-4}$ on hardware, however, takes about a million logical cycles per latency and distance.
\Dynamite{} (Opt-1+Opt-2) needs even more logical cycles: its safe thresholds, the per-round limits on the syndrome vector HW, require the logical failure rate for each value of the HW.
\Dynamite{} therefore profiles only the measurement error at each latency, measured or from Equation~\ref{eq:meas_error}, and obtains $L_{opt1}$ and the safe thresholds $\tau_r$ from these errors through simulation; hardware QEC experiments at a few latencies near this $L_{opt1}$ can then confirm the selection.}

\section{Discussion}
\label{sec:discussion}

\begin{figure}[htp]
    \centering
     \includegraphics[width=\columnwidth]{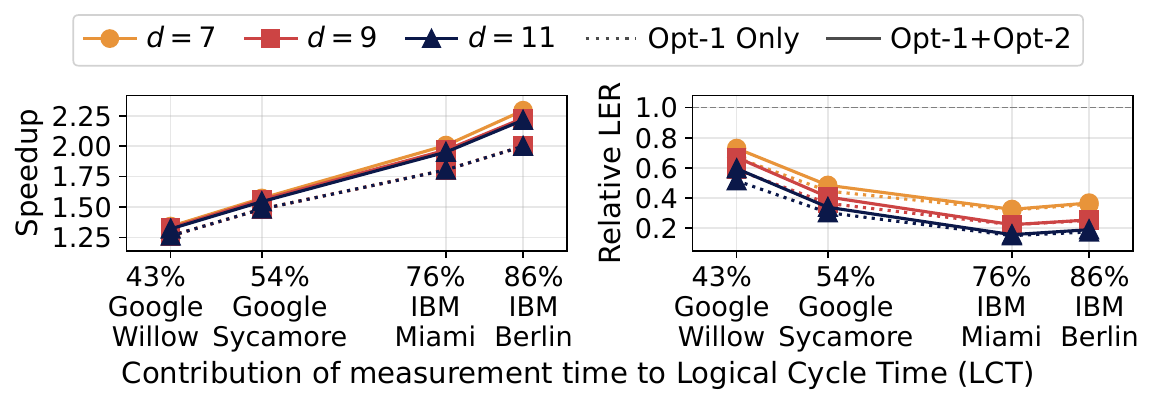}
     \vspace{-0.25in}
     \caption{{Speedup and relative LER of \Dynamite{} versus the contribution of measurement time to LCT ($p{=}0.1\%$, 100 million logical cycles).}}
    \label{fig:regime}
\end{figure}

\noindent\review{\textbf{Regime of usefulness:} 
\Dynamite{} speedup depends on contribution of measurement time to logical cycle time (LCT). Figure~\ref{fig:regime} shows performance of four devices with measurements taking 43 to 86\% of LCT: speedup grows with this contribution from 1.3$\times$ on Willow to 2.3$\times$ on Berlin. Relative LER improvements are higher for lower values of $T_1/T_2$ due to better suppression of decoherence errors (Sycamore outperforms Willow and Miami outperforms Berlin).}

\noindent\textbf{Impact on real-time decoding:} Faster measurements shorten the syndrome-round and tighten the budget that decoders must meet for real-time correction. A wide range of hardware-accelerated decoders (LUT-based~\cite{das2022lilliput}, MWPM-based~\cite{wu2023fusion, vittal2023astrea}, predecoder~\cite{alavisamani2024promatch}, neural~\cite{overwater2022neural}, cryogenic~\cite{ueno2021qecool}, and FPGA/ASIC~\cite{barber2025collision, wu2025micro, liyanage2023scalable}) can pair with \Dynamite{} to meet this tighter budget. Moreover, average decoding latency is very low~\cite{Higgott2025sparseblossom}, with high latencies restricted to the rare cycles with many errors, so a decoder rarely exceeds the real-time budget.

\noindent\textbf{Compatibility with other platforms:} Although focused on superconducting qubits, \Dynamite{} generalizes to other platforms where measurement is a bottleneck. Neutral atoms spend ${\sim}75\%$ of the QEC cycle on fluorescence imaging (1--10\,ms, orders of magnitude slower than gates~\cite{wintersperger2023neutral, mude2025enabling}), showing similar latency--fidelity trade-offs. Adapting \Dynamite{} to neutral atoms or trapped ions requires only LLT/EBT re-profiling; relaxed latency constraints enable software implementation.
% trapped-ion measurement latency: 120us (TrapSimd)
% -> intra-trap qubit shift, swap is critical

\noindent\textbf{Finer-grained measurements:} \Dynamite{} selects from a discrete latency set (for example, 50\,ns steps on Google Willow). A finer grid gives the online policy more operating points and converts more unused budget into speedup, requiring only finer calibration and no change to the scheduling hardware.

\section{Related Work}

%We describe related work and compare and contrast as appropriate. 
%We describe related work, drawing comparisons to our approach where relevant.

\noindent\textbf{Adaptive QEC:}   
Berthusen et al.~\cite{berthusen2025adaptive} propose adaptive syndrome extraction that measures only those stabilizers that are likely to yield useful information, reducing the number of CNOTs and physical qubits needed.
Tansuwannont et al.~\cite{tansuwannont2023adaptive} introduce adaptive measurement rounds for Shor-style QEC to reduce the worst-case number of syndrome rounds using intermediate results. \Dynamite{} is orthogonal to these. Rather than changing \textit{which} or \textit{how many} stabilizers to measure, \Dynamite{} dynamically adjusts the duration of each measurement. 

\noindent\textbf{Device-level optimizations:} The measurement latency--fidelity tradeoff has been extensively characterized at the qubit level: Bengtsson et al.~\cite{bengtsson2024model} achieve $1.5\%$ error per qubit with $500$\,ns readout, and Chen et al.~\cite{chen2023transmon} reach $99.5\%$ single-shot fidelity within $140$\,ns. These optimizations face scalability challenges; \Dynamite{} instead schedules QEC cycles around this trade-off, remaining effective as devices improve. 

\noindent\textbf{QEC-specific measurement:} Pattison et al.~\cite{pattison2021_softInfo} use continuous readout outcomes to improve thresholds by 25\% and model the resolution--disturbance trade-off; Heussen et al.~\cite{heussen2024measurement} replace readouts with coherent operations for measurement-free QEC. \Dynamite{} is complementary, dynamically tuning measurement time per round to balance fidelity and speed.

\section{Conclusion}
\label{sec:conclusion}

%Qubit measurements account for 43--86\% of logical cycle times (LCTs) and dictate program latencies. 
Qubit measurements account for 43--86\% of logical cycle times (LCTs) and dictate program latencies. We propose \Dynamite{}, a lightweight real-time measurement latency selection policy built on two insights. \textit{First}, rather than relying on calibrated latencies from device providers that only maximize readout fidelity, \Dynamite{} selects latencies that optimize the logical performance collectively. \textit{Second}, not all logical cycles are equal; most encounter far fewer errors than what can be tolerated (\textit{error budget}) by the quantum error correction (QEC) code and its decoder. \Dynamite{} effectively uses the error budget by dynamically using faster measurements when its consumption is low. While this raises the error rate of the measurements, \Dynamite{} ensures that the logical error rate (LER) is not compromised by tracking the error budget in real-time and ensuring that it is not exceeded. Overall, \Dynamite{} improves LER and LCT by 0.46$\times$ and 1.69$\times$ on average and by up to 0.13$\times$ and 2.29$\times$, respectively, compared to baseline.

\section*{Acknowledgements}
We thank Blake Johnson from IBM Quantum for the insightful technical discussions surrounding dynamic measurements, synchronization mechanisms for logical operations, and complex readout error modeling by accounting for resonator characteristics.
We also thank the generous financial support from the Cockrell School of Engineering and the AMD endowment at the University of Texas at Austin.    

%%%%%%% -- PAPER CONTENT ENDS -- %%%%%%%%

%%%%%%%%% -- BIB STYLE AND FILE -- %%%%%%%%
\bibliographystyle{IEEEtran} % references [1, 2, 3, ...] -> [1-3]
\bibliography{refs}
%%%%%%%%%%%%%%%%%%%%%%%%%%%%%%%%%%%%

\end{document}